\documentclass[aps,pra,reprint, longbibliography, twocolumn,floatfix,superscriptaddress]{revtex4-1}

\usepackage[T1]{fontenc}
\usepackage{lmodern}
\usepackage[utf8]{inputenc}
\usepackage[english]{babel}

\usepackage{microtype}
\usepackage{indentfirst}  

\usepackage{amsthm, amssymb, amstext, amsmath, bbm}
\usepackage{dsfont}

\usepackage{float}
\usepackage{tikz}

\usepackage{epstopdf}
\usepackage{blkarray} 
\usepackage{MnSymbol}

\usepackage{soul}
\usepackage{braket}

\usepackage{xcolor}

\usepackage{hhline}

\usepackage{enumitem}

\definecolor{darkred}{rgb}{0.5,0,0}
\definecolor{darkgreen}{rgb}{0,0.5,0}
\definecolor{darkblue}{rgb}{0,0,0.5}

\usepackage[colorlinks]{hyperref}
\hypersetup{colorlinks, linkcolor=darkred, filecolor=darkgreen,
urlcolor=darkblue, citecolor=darkgreen}

\newcommand{\LL}{\mathcal{L}}
\newcommand{\DD}{\mathcal{D}}

\newcommand{\rhot}{\hat{\rho}(t)}
\newcommand{\de}{{\rm d}}
\newcommand{\sss}{\hat{\rho}_{\rm ss}}
\newcommand{\eig}[1]{\hat{\rho}_{#1}}

\renewcommand{\Re}[1]{\mathbb{R}\mathrm{e}\left[#1\right]}
\renewcommand{\Im}[1]{\mathbb{I}\mathrm{m}\left[#1\right]}

\newcommand{\supmat}[1]{\bar{\bar{#1}}}

\newcommand{\abs}[1]{\lvert #1 \rvert}

\newcommand{\Tr}[1]{\mathrm{Tr}\!\left[#1\right]}
\usepackage{epstopdf}

\newtheorem{lemma}{Lemma}

\newtheorem{theorem}{Theorem}

\makeatletter
\newcommand*\bigcdot{\mathpalette\bigcdot@{.5}}
\newcommand*\bigcdot@[2]{\mathbin{\vcenter{\hbox{\scalebox{#2}{$\m@th#1\bullet$}}}}}
\makeatother

\begin{document}
 
\title{Quantum exceptional points of non-Hermitian Hamiltonians and Liouvillians:\\  The effects of quantum jumps}

\author{Fabrizio Minganti}\email{fabrizio.minganti@riken.jp }
\affiliation{Theoretical Quantum Physics Laboratory, RIKEN Cluster
	for Pioneering Research, Wako-shi, Saitama 351-0198, Japan}

\author{Adam Miranowicz}\email{adam@riken.jp}
\affiliation{Theoretical Quantum Physics Laboratory, RIKEN Cluster
	for Pioneering Research, Wako-shi, Saitama 351-0198, Japan}
\affiliation{Faculty of Physics, Adam Mickiewicz University,
	PL-61-614 Poznan, Poland}

\author{Ravindra W. Chhajlany}\email{ravi@amu.edu.pl}
\affiliation{Theoretical Quantum Physics Laboratory, RIKEN Cluster
	for Pioneering Research, Wako-shi, Saitama 351-0198, Japan}
\affiliation{Faculty of Physics, Adam Mickiewicz University,
	PL-61-614 Poznan, Poland}

\author{Franco Nori}\email{fnori@riken.jp}
\affiliation{Theoretical Quantum Physics Laboratory, RIKEN Cluster
	for Pioneering Research, Wako-shi, Saitama 351-0198, Japan}
\affiliation{Physics Department, The University of Michigan, Ann
	Arbor, Michigan 48109-1040, USA}

\date{\today}

\begin{abstract}
Exceptional points (EPs) correspond to degeneracies of open
systems. These are attracting much interest in optics, optoelectronics, plasmonics, and condensed matter physics. In the classical and semiclassical approaches, Hamiltonian EPs (HEPs) are usually defined as degeneracies of non-Hermitian Hamiltonians such that at least two eigenfrequencies are identical and the corresponding eigenstates coalesce. HEPs result from continuous, mostly slow, nonunitary evolution without quantum jumps. Clearly, quantum jumps should be included in a fully quantum approach to make it equivalent to, e.g., the
Lindblad master-equation approach. Thus, we suggest to define EPs
via degeneracies of a Liouvillian superoperator (including the
full Lindbladian term, LEPs), and we clarify the relations between HEPs and LEPs. We prove two main theorems: Theorem~1 proves that, in the quantum limit, LEPs and HEPs must have essentially different properties. Theorem~2 dictates a condition under which, in the ``semiclassical'' limit, LEPs and HEPs recover
the same properties. In particular, we show the validity of Theorem 1
studying systems which have (1) an LEP but no HEPs, and (2)
both LEPs and HEPs but for shifted parameters. As for Theorem 2,
(3) we show that these two types of EPs become essentially
equivalent in the semiclassical limit. We introduce a series of mathematical techniques to unveil
analogies and differences between the HEPs and LEPs.
We analytically compare LEPs
and HEPs for some quantum and semiclassical prototype models with
loss and gain.
\end{abstract}

\maketitle

\tableofcontents

\section{Introduction}

Exceptional points (EPs) are attracting increasing interest, both theoretically and experimentally, in diverse fields including
optics, condensed-matter physics, plasmonics, and even electronics. For example, as summarized in very recent
reviews~\cite{Ozdemir2019,Miri2019}, EPs are considered for novel enhanced sensing apparatus and are relevant to describe dynamical phase transitions and in the characterization of topological phases of matter in open systems.
This research in EPs was triggered two decades ago by the introduction of non-Hermitian quantum mechanics \cite{Bender1998} or, more specifically, by the discovery of non-Hermitian Hamiltonians (NHHs) with real eigenvalues for parity-time-symmetric nonconservative systems \cite{ChristodoulidesBOOK}. An EP of an NHH (which for short we refer to as a Hamiltonian EP or an HEP) refers to the NHH degeneracies where two (or more) eigenfrequencies coincide and the corresponding eigenstates coalesce. Since an EP corresponds to a nondiagonalizable operator, standard Hermitian Hamiltonians cannot display any EP. It is the nonunitary effect of the environment that induces the emergence of EPs. Such points can be found, e.g., by balancing the attenuation, amplification, gain saturation, as well as various Hamiltonian coupling strengths of an open system (as experimentally shown in, e.g., Refs.~\cite{PengScience14,GaoNature15,NaghilooNatPhys19}).

The dynamics of an open \emph{quantum} system is characterized by the
presence of dissipative terms describing the progressive loss of
energy, coherence, and information into the environment. Under very
general hypotheses, the equation of motion can be captured by a Lindblad form consisting of a Hermitian Hamiltonian part, describing the coherent evolution of the system, and a non-Hermitian part, the so-called Lindblad dissipators. These Lindblad dissipators admit a fascinating interpretation in terms of quantum maps and measurement theory \cite{Wiseman_BOOK_Quantum,Haroche_BOOK_Quantum,Barnett_BOOK_Info,ParisEPJST2012}, and can be divided into two parts: The first one represents a \emph{coherent nonunitary dissipation} of the system, transforming the Hamiltonian in an NHH. 
The second one describes \emph{quantum jumps}, which are the effect of a continuous measurement performed by the environment on the system.

The instantaneous switching between energy levels in quantum systems, caused by quantum jumps, is pivotal to correctly describe microscopic open systems. These quantum jumps lay at the foundation of quantum physics, being necessary to obtain a consistent measurement theory once the environment is taken into account \cite{Wiseman_BOOK_Quantum,DaleyAdvancesinPhysics2014,Carmichael_BOOK_1,Carmichael_BOOK_2}. Quantum jumps have been observed in countless experiments, including ionic \cite{NagourneyPRL86,SauterPRL96,BergquistPRL99}, atomic
\cite{PeilPRL99,BascheNat95,GleyzesNat07,GuerlinNat07,DelegliseNat08,SayrinNat11}, solid-state \cite{JelezkoAPL02,NeumannScience10,RobledoNat11}, and
superconducting circuit setups \cite{VijayPRL11,HatridgeScience13,SunNature2014,OfekNat16,Minev2019}.
Hence, to correctly describe exceptional points of quantum systems, one
\emph{must} consider quantum jumps.

The vast majority of studies on EPs, especially in the context
of parity-time-symmetric systems, have been limited to classical
or semiclassical models where quantum jumps were ignored (e.g., \cite{JingPRL14,PengScience14}). Indeed, the standard calculation of EPs (i.e., HEPs) is based on finding degeneracies of non-Hermitian Hamiltonians. Thus, such approaches cannot be completely equivalent to the standard Lindblad master equation, as clearly seen by referring to the quantum trajectory method (also known as the quantum jump method)~\cite{Haroche_BOOK_Quantum,PlenioRMP98,CarmichaelPRL93,DalibardPRL92,MolmerJOSAB93}.

The time evolution of a system obeying a Lindblad master equation is captured by a Liouvillian superoperator.
Since the Liouvillian is a non-Hermitian matrix, it too can exhibit EPs \cite{LidarLectureNotes,  AlbertPRA14,MingantPRA18_Spectral,MacieszczakPRL16,SarandyPRA05_Adiabatic,HatanoMP19,ProsenJST10,MoosSciPost19}.
Liouvillian EPs (LEPs) are defined via degeneracies of
Liouvillians (including the full Lindbladian terms), i.e., when two
(or more) eigenfrequencies and the corresponding eigenstates of a
given Liouvillian coalesce.
The physical meaning of LEPs and its relation to HEPs, however, are crucial to correctly understand HEPs in the quantum case.

The main objective of this paper is to point out the similarities and differences between HEPs and LEPs.
To do that, we introduce the Liouvillian without quantum jumps $\LL'$, as well as the exceptional points resulting from a Liouvillian without quantum jumps (LEP's).
We prove the severe limits of the NHH approach to correctly capture the full quantum regime, and we demonstrate how quantum jumps can also affect the semiclassical dynamics of a system. 
In this  regard, we provide a procedure to generalize the semiclassical
HEPs to include quantum jumps. 
We prove theorems about the general properties of HEPs and LEPs, showing their equivalence in the semiclassical regime and some fundamental differences in the
quantum regime. 
We also compare the basic properties of HEPs and LEPs. 
By introducing a series of mathematical techniques, we unveil analogies and differences between the HEPs and LEPs. 
We demonstrate these similarities and discrepancies on three prototype examples.

This paper is organized as follows: In
Sec.~\ref{Sec:NHH_L_semiclassical}, we discuss the semiclassical limit, how in this limit the NHH stems from a Liouvillian and how, vice versa, a Liouvillian is the minimal extension of an NHH to the quantum regime. 
In Sec.~\ref{Sec:Spectrum} we provide the main results of this paper, i.e., Theorems~\ref{Thm:no_HEP_are_LEP}~and~\ref{Thorem:When_NHH_is_true_for_EPs}, which prove some relations between the spectra of an NHH and the corresponding Liouvillian. 
In Secs.~\ref{Sec:Example_1},~\ref{Sec:Example_2}, and~\ref{Sec:Example_3}, we demonstrate the validity of the
theorems on three examples.
Finally, in the Appendix~\ref{Sec:superoperators} we recall, for pedagogical reasons, some useful properties of superoperators.

In the main article we will use all the previously introduced abbreviations. 
In Table \ref{Tab:Abb} we concisely list them to facilitate the reading of the article. 

\begin{table}
	\begin{tabular}{c|c}
	Full Name & Abbreviation \\
	\hhline{=|=}
	Non-Hermitian Hamiltonian     &  NHH\\
	\hline
	Liouvillian     & $\LL$ \\
	\hline
	Liouvillian without quantum jumps     & $\LL'$ \\
	\hline
	Exceptional point & EP \\
	\hline
	Hamiltonian exceptional point & HEP
	\\
	\hline
	Liouvillian exceptional point & LEP \\
	\hline LEP without quantum jumps & LEP'
\end{tabular}
\caption{List of abbreviation used in the main article, and corresponding full names.}
\label{Tab:Abb}
\end{table}

\section{Non-Hermitian Hamiltonians, Liouvillians, and their semiclassical approximation}
\label{Sec:NHH_L_semiclassical}

In this section, we introduce in detail the concept of non-Hermitian Hamiltonian (NHH) and Liouvillian stemming from the Lindblad master equation.
In particular, we explain that the semiclassical limit of a
Liouvillian is an NHH, and we provide a physical interpretation
of the resulting effective Hamiltonian. 
Vice versa, we demonstrate that the Lindblad master equation is a minimal quantum map that extends the behavior of a non-Hermitian Hamiltonian to its ``quantum'' limit.

Before proceeding further, let us clarify the usage of the term
\emph{semiclassical limit} to be applied in this paper. In the
literature, the ``semiclassical approximation'' is loosely and
widely used with different meanings~\cite{Arndt2009}. Thus, the
semiclassical regime can be defined in various ways depending on
its physical context~\cite{SchlosshauerBook}. These meanings
include:\\
\textbf{(1)} In the traditional interpretation of nonrelativistic quantum
mechanics, the semiclassical limit corresponds to assuming
$\hbar\to 0$, transforming operators into variables, and replacing
the Hilbert space tensor-product structure with the direct sum of
classical phase spaces.\\
\textbf{(2)} One can also refer to the semiclassical regime of a quantum
system that can be well approximated by a classical model for high
quantum numbers. A classical example can be provided by the
coherent-state approximation of the electromagnetic field in
quantum optics~\cite{Walls_BOOK_quantum}, where the evolution of a
state inside a cavity is well captured by the evolution of a
complex number. Moreover, this is often the case when discussing
room-temperature condensed-matter systems for which quantum
characteristics are well captured by phenomenological classical
theories (e.g., the Drude scattering theory for electrons or the
Johnson-Nyquist noise).\\
\textbf{(3)} Another meaning of the semiclassical approximation regards 
composite systems, that is, when the system can be
described as a classical subsystem interacting with a quantum one.
For example, the standard optical Bloch
equations~\cite{Haroche_BOOK_Quantum, Carmichael_BOOK_1} describe
a quantum two-level system coupled to a classical electromagnetic
field.\\
\textbf{(4)} Moreover, one can consider that, during the dissipative evolution of a
quantum system, different physical properties evolve at different rates, dictating a 
passage from the quantum to the semiclassical regime. In particular, one can introduce 
the pointer states of dissipation as the classical states emerging from a
prolonged interaction with a complex environment~\cite{ZurekRMP03}.\\
\textbf{(5)} In the literature about EPs, the effects of quantum noise are
neglected by claiming that the model is semiclassical. Similarly,
in the present discussion, the semiclassical limit means that we neglect 
the action of quantum jumps without taking much care of which of the previous 
four criteria can be applied.

Specifically, we will have a well-defined semiclassical regime (or 
semiclassical limit) of the Markovian dynamics of a given quantum
system if all (or at least some nontrivial) EPs of a non-Hermitian
Hamiltonian are effectively the same as those of a corresponding
Liouvillian in a Lindblad master equation \emph{with} quantum
jumps terms. We stress that all the EPs of a non-Hermitian
Hamiltonian (i.e., HEPs) are exactly the same as the EPs of 
a Liouvillian \emph{without} quantum jump terms (LEP's).

\subsection{The Liouvillian and its semiclassical limit}

The time evolution of an open quantum system weakly interacting
with a Markovian (i.e., memoryless) environment can be expressed
using the so-called Lindblad master equation
\cite{Carmichael_BOOK_1,BreuerBookOpen,Haroche_BOOK_Quantum,Walls_BOOK_quantum,Gardiner_BOOK_Quantum,Wiseman_BOOK_Quantum}
(hereafter, we set $\hbar=1$):
\begin{equation}\label{Eq:Lindblad}
\frac{\partial \rhot}{\partial t} = \LL \rhot =- i \left[\hat{H},
\rhot\right] + \sum_\mu \DD[\hat{\Gamma}_\mu] \rhot,
\end{equation}
where $\rhot$ is the density matrix of a system at a time $t$ and
$\DD[\hat{\Gamma}_\mu]$ are the dissipators associated with the
jump operators $\hat{\Gamma}_\mu$, while $\LL$ is the so-called
Liouvillian superoperator (for a detailed discussion about
superoperators, see Appendix~\ref{Sec:superoperators}). The
density matrix $\rhot$ can describe both pure states $\ket{\phi}\bra{\phi}$ and probabilistic mixture
$\sum_i p_i \ket{\phi_i}\bra{\phi_i}$\cite{Cohen-Tannoudji_BOOK_Quantum_Vol_1,Carmichael_BOOK_1,Gardiner_BOOK_Quantum,Haroche_BOOK_Quantum,ParisEPJST2012,RivasBOOK_Open}.
Each dissipator is defined by the Lindbladian
\begin{equation}\label{Eq:Dissipator}
\DD [\hat{\Gamma}_\mu] \rhot = \hat{\Gamma}_\mu \rhot
\hat{\Gamma}_\mu^\dagger - \frac{\hat{\Gamma}_\mu^\dagger
	\hat{\Gamma}_\mu}{2} \rhot - \rhot \frac{\hat{\Gamma}_\mu^\dagger
	\hat{\Gamma}_\mu}{2}.
\end{equation}
The Lindblad master equation admits a very appealing interpretation
as the time evolution of a system which is continuously monitored
by an environment \cite{Haroche_BOOK_Quantum}. In this regard, the
effect of $\DD [\hat{\Gamma}_\mu]$ on the density matrix $\rhot$
can be split into two parts \cite{Wiseman_BOOK_Quantum}: the
continuous nonunitary dissipation terms, $\hat{\Gamma}_\mu^\dagger
\hat{\Gamma}_\mu \rhot + \rhot \hat{\Gamma}_\mu^\dagger
\hat{\Gamma}_\mu$, and the \emph{quantum jump} terms,
\begin{equation}
\mathcal{J}[\hat{\Gamma}_\mu]\rhot=\hat{\Gamma}_\mu \rhot
\hat{\Gamma}_\mu^\dagger.
\end{equation} 
The dissipation describes the
continuous losses of energy, information, and coherence of the
system into the environment, while the quantum jumps describe the
effect of the measurement on the state of the system
\cite{Barnett_BOOK_Info,Wiseman_BOOK_Quantum,Haroche_BOOK_Quantum}.
We label the term $\mathcal{J}[\hat{\Gamma}_\mu]$ a quantum jump
since in a quantum trajectory approach (i.e., a wave-function Monte
Carlo method)
\cite{MolmerJOSAB93,DalibardPRL92,DaleyAdvancesinPhysics2014,CarmichaelPRL93},
those are the terms responsible for the abrupt stochastic change
of the wave function. In this regard, given a Lindblad master
equation describing the microscopic physics of a given system, it
is easy to obtain the corresponding ``semiclassical limit'' by
neglecting the effect of quantum jumps, and introducing an
effective NHH of the form
\begin{equation}
\hat{H}_{\rm eff} = \hat{H} - i \sum \hat{\Gamma}_\mu^\dagger \hat{\Gamma}_\mu/2.
\end{equation}
An equation of motion for a generic density matrix $\rhot$ thus
becomes:
\begin{equation}\label{Eq:Linblad_semiclassical}
\frac{\partial \hat{\rho}(t)}{\partial t} = \LL'  \hat{\rho}(t)= - i \left(\hat{H}_{\rm eff}\hat{\rho}(t) -  \hat{\rho}(t) \hat{H}_{\rm eff}^\dagger \right),
\end{equation}
where we have introduced the Liouvillian without quantum jumps
$\LL'$. Indeed, in this evolution, one assumes that the effect of the jump operators $\hat{\Gamma}_\mu$ 
is negligible due to the semiclassical nature of the system state.
We stress that this condition requires that not only the steady state but also all the states explored by the Liouvillian dynamics can be well described via a non-Hermitian Hamiltonian.
Indeed, as it will be detailed in Sec.~\ref{sec:nogo}, the LEPs capture the dynamical properties of a system relaxing towards its steady state and as such cannot be captured by the properties of the steady state alone.

We note that the master equation in Eq.~\eqref{Eq:Lindblad} can be rewritten
in terms of $\hat{H}_{\rm eff}$ as follows:
\begin{equation}\label{Eq:Lindblad2}
\frac{\partial \rhot}{\partial t} = \mathcal{L} \rhot =- i
\left(\hat{H}_{\rm eff}\hat{\rho}(t) -  \hat{\rho}(t) \hat{H}_{\rm eff}^\dagger \right)+ \sum_\mu \hat{\Gamma}_\mu
\rhot \hat{\Gamma}_\mu^\dagger.
\end{equation}
Thus, it is clear that given a non-Hermitian Hamiltonian
$\hat{H}_{\rm eff}$, together with the quantum jump terms
$\hat{\Gamma}_\mu \rhot \hat{\Gamma}_\mu^\dagger$
in Eq.~(\ref{Eq:Lindblad2}), one can fully describe the quantum dynamics
of a dissipative system within the Lindblad formalism. However, a
natural way to calculate the EPs of a non-Hermitian Hamiltonian
with the quantum jump terms requires the use of a superoperator
rather than the operator $\hat{H}_{\rm eff}$ (see also Appendix~\ref{Sec:superoperators}). This is because the
quantum jump operators are on the left- and right-hand sides of a
density matrix $\rhot$ in the term $\hat{\Gamma}_\mu \rhot
\hat{\Gamma}_\mu^\dagger$. Such a superoperator is actually the
Liouvillian $\mathcal{L}$ studied here.

Finally, $\hat{H}_{\rm eff}$ naturally emerges when discussing \emph{quantum trajectories} and \emph{postselection}. 
Quantum trajectories describe a system whose environment is continuously and perfectly probed \cite{MolmerJOSAB93,Haroche_BOOK_Quantum,Carmichael_BOOK_2,Wiseman_BOOK_Quantum,DaleyAdvancesinPhysics2014}.
In this formalism, the system state is captured by a stochastic wave function $\ket{\psi(t)}$. 
In a counting trajectory, when no quantum jump happens, $\ket{\psi(t)}$ evolves smoothly according to $\hat{H}_{\rm eff}$. 
Otherwise, the wave function abruptly changes under the action of a jump operator, $\hat{\Gamma}_\mu \ket{\psi(t)}$ \cite{Wiseman_BOOK_Quantum}.
The Lindblad master equation describes the average over infinite quantum trajectories, i.e., infinitely many experiments. 
In some of these experiments, a quantum jump took place.
If, instead, one considers only those trajectories where no quantum jumps happened, the average over many different trajectories would be described by $\LL'$.
In this regard, to access the NHH in its quantum limit, one can use post-selection techniques, as done in Ref.~\cite{NaghilooNatPhys19}.

\subsection{Making sense of non-Hermitian Hamiltonians \\ in the quantum limit}

We try now to reconcile the concept of NHHs with that of a
quantum map.

Let us consider the following NHH:
\begin{equation}
\hat{H}_{\rm eff}= \hat{H} + \hat{A},
\end{equation}
where we introduce the Hermitian operator $\hat{H}= ( \hat{H}_{\rm
	eff}+\hat{H}^\dagger_{\rm eff})/2$ and the anti-Hermitian one
$\hat{A}=(\hat{H}_{\rm eff}-\hat{H}^\dagger_{\rm eff} )/2$. One
can prove that the most general form of a \emph{linear,
	Hermiticity- and trace-preserving, and completely positive} quantum
map describing the time evolution of the density matrix $\rhot$ is
a superoperator $\mathcal{M}$
\cite{Haroche_BOOK_Quantum,Barnett_BOOK_Info,ParisEPJST2012}, defined by
\begin{equation}\label{Eq:Most_General}
\hat{\rho}(t+\tau) =\mathcal{M} \rhot= \sum_{\mu} \hat{M}_\mu
\rhot \hat{M}_\mu^\dagger \quad {\rm and } \quad \sum_{\mu}
\hat{M}^\dagger_\mu \hat{M}_\mu = \mathds{1},
\end{equation}
where $\mathcal{M}$ are the Kraus operators. Since this NHH captures well the dynamics of the system in its semiclassical limit,
the time evolution of a generic density matrix $\rhot$ under such
an NHH is
\begin{equation}\label{Eq:Evolution_non_hermitian}
\begin{split}
\hat{\rho}(t+\tau) &=  \rhot - i \tau \left( \hat{H}_{\rm eff}
\rhot - \rhot \hat{H}^\dagger_{\rm eff}
\right) + \mathcal{R} \rhot \\
& = \rhot - i \tau \left[\hat{H}, \rhot \right] -  \tau \left\{i
\hat{A}, \rhot \right\} + \mathcal{R} \rhot,
\end{split}
\end{equation}
where the superoperator $\mathcal{R}$ is the additional term
needed to recover a Kraus map, while $\left[\bigcdot,
\bigcdot\right]$ and $\left\{\bigcdot, \bigcdot\right\}$ represent
the commutator and anticommutator, respectively.

Since semiclassically the density operators evolve smoothly under the action of $\hat{H}_{\rm eff}$, we assume that in the quantum limit $\rhot$ evolves according to
Eq.~\eqref{Eq:Most_General} as
\begin{equation}
\label{Eq:Kraus_map_Lindblad} \hat{\rho}(t+\tau) = \mathcal{M}
\rhot = \sum_{\mu} \hat{M}_\mu^\dagger \rhot \hat{M}_\mu = \rhot +
\tau \frac{\de \rhot}{\de t} + \mathcal{O}(\tau^2).
\end{equation}
To identify the form of $\mathcal{R}$, we note that $\mathcal{R}
\rhot \ll \tau \left\{i \hat{A}, \rhot \right\}$ holds in the
semiclassical limit. That is, the terms stemming from
$\mathcal{R}$ in a semiclassical picture produce only a constant
shift, plus terms which are small compared to the action of the
non-Hermitian part of the NHH. 
Hence, by comparing
Eqs.~\eqref{Eq:Linblad_semiclassical},~\eqref{Eq:Evolution_non_hermitian},~and~
\eqref{Eq:Kraus_map_Lindblad}, we deduce that
\begin{equation}
\hat{M}_0 = \mathds{1} - i \tau \hat{H}_{\rm eff},
\end{equation}
so that
\begin{equation}
\hat{M}_0 \rhot \hat{M}_0^\dagger =\rhot  - i \tau \left( \hat{H}_{\rm eff} \rhot - \rhot \hat{H}^\dagger_{\rm eff}
\right) + \mathcal{O}(\tau^2).
\end{equation}
Therefore, we conclude that
\begin{equation}
\mathcal{R} \rhot= \sum_{\mu \neq 0} \hat{M}_\mu^\dagger \rhot
\hat{M}_\mu.
\end{equation}
If we assume that there is only another Kraus operator $\hat{M}_1$,
i.e., $\mathcal{R}=\hat{M}_1^\dagger \bigcdot\hat{M}_1$, to
satisfy Eq.~\eqref{Eq:Most_General} we have
\begin{equation}
\begin{split}
\hat{M}_1^\dagger \hat{M}_1 &=\mathds{1} - \hat{M}_0^\dagger \hat{M}_0 = -i \tau (\hat{H}_{\rm eff} - \hat{H}_{\rm eff}^\dagger) \\
&  -i \tau (\hat{H}_{\rm eff} - \hat{H}_{\rm eff}^\dagger) = - 2 i
\tau A.
\end{split}
\end{equation}
From this relation, we define $\hat{M}_1=\Gamma=\sqrt{-2 i A}$ and
obtain
\begin{equation}\label{Eq:Lindblad_for_NHH_incremental}
\begin{split}
\hat{\rho}(t+\tau) &= \rhot - i \tau \left[\hat{H}, \rhot \right] - \tau \left\{\frac{\hat{\Gamma}^\dagger \hat{\Gamma}}{2}, \rhot \right\} + \hat{\Gamma}^\dagger \rhot \hat{\Gamma} \\
&= \rhot - i \tau \left[\hat{H}, \rhot \right] + \tau
\DD[\hat{\Gamma}] \rhot,
\end{split}
\end{equation}
where $\DD[\hat{\Gamma}]=\hat{\Gamma} \bigcdot
\hat{\Gamma}^\dagger - \frac{1}{2} \left\{\hat{\Gamma}^\dagger
\hat{\Gamma}, \bigcdot \right\}$ is the standard Lindbladian
dissipator of Eq.~\eqref{Eq:Dissipator}. 
Indeed, we have proved that the minimal additional
term necessary to properly extend an NHH to its quantum regime is
the jump superoperator $\mathcal{J}[\hat{\Gamma}]$.

To conclude, we recast Eq.~\eqref{Eq:Lindblad_for_NHH_incremental}
in its differential form, obtaining the Lindblad master equation,
\begin{equation}\label{Eq:Liouvillian}
\frac{\partial \rhot}{\partial t} = - i \left[\hat{H},
\rhot\right] + \DD[\hat{\Gamma}] \rhot \equiv \LL \rhot.
\end{equation}
We stress that, this minimal model (exploiting only two
Kraus operators to describe the system) may not be sufficient to
fully capture the physics underlying the NHH.
However, it is the simplest generalization allowing one to study open quantum systems far from the semiclassical regime.

\emph{The properties of} $\LL$, $\LL'$, \emph{and}  $\hat{H}_{\rm
	eff}$ \emph{will be the focus of this article}.

\section{Liouvillian spectrum, exceptional points, and their physical meaning in open quantum systems}
\label{Sec:Spectrum}

We introduce the eigenvalues and eigenvectors of $\hat{H}_{\rm
	eff}$ via the relation
\begin{equation}
\hat{H}_{\rm eff} \ket{\phi_i} = h_i \ket{\phi_i}.
\end{equation}
In the same way in which we diagonalize a $\hat{H}_{\rm eff}$, one
can obtain much information about an open quantum system by
studying the spectrum of a Liouvillian. For a time-independent
Liouvillian, there always exists at least one steady state (if the dimension of the Hilbert space is finite
\cite{RivasBOOK_Open,BreuerBookOpen}), i.e., a matrix which does
not evolve under the Lindblad master equation. Such a steady state
$\sss$ is an eigenmatrix of a given Liouvillian, since
\begin{equation}
\LL \sss = 0.
\end{equation}
In this regard, the steady state plays a similar role to the
ground state of a Hamiltonian.

Even if the steady state has a privileged role, its knowledge is
not enough to fully determine all the properties of the system.
Indeed, many interesting phenomena can occur in the dynamics
towards a steady state. Therefore, one has to study the spectrum
of the Liouvillian superoperator $\LL$, whose eigenmatrices  and
eigenvalues are defined via the relation
\begin{equation}\label{eq:Liouvillian_Eigenstates}
\mathcal{L} \eig{i} = \lambda_i \eig{i}.
\end{equation}
By introducing the Hilbert-Schmidt inner product (see Appendix~\ref{Sec:superoperators}),
one can normalize the eigenmatrices $|\eig{i}|^2 = \Tr{\eig{i}^\dagger \eig{i}}=1$.
Since a Liouvillian does not need to be a Hermitian superoperator, in general different eigenmatrices
will not be orthogonal:
$\Tr{\eig{i}^\dagger \eig{j}}\neq \delta_{i, \, j}$.
Moreover, $\LL$ admits both left and right eigenmatrices, where the former are
defined by
\begin{equation}\label{eq:Liouvillian_left_Eigenstates}
\mathcal{L}^\dagger \hat{\sigma}_i = \lambda_i^* \hat{\sigma}_i.
\end{equation}
The left and right eigenmatrices are mutually orthonormal in the
sense that one can always normalize them in such a way that $\Tr{\hat{\sigma}_i
	\eig{j}}=\delta_{i, \, j}$. 
Therefore, if a Liouvillian is
diagonalizable (that is, apart from the LEPs
\cite{MacieszczakPRL16}), any density matrix $\rhot$ can be
written as
\begin{equation}
\label{Eq:eigenvectordecomposition} \rhot=\sum_i c_i(t) \eig{i},
\end{equation}
where $c_i(t)=\exp{(\lambda_i t)} \Tr{\hat{\sigma}_i
	\hat{\rho}(0)}$ (see Lemma~\ref{Lemma:Evolution_Eigenstate} below). 
It can be proved
\cite{BreuerBookOpen,RivasBOOK_Open} that $ \forall i$,
$\Re{\lambda_i}\leq 0$. Therefore, the real part of the
eigenvalues $\lambda_i$ are responsible for the relaxation rate of
any expectation value towards the steady state. For convenience,
we sort the eigenvalues in such a way that
$\abs{\Re{\lambda_0}}<\abs{\Re{\lambda_1}} < \ldots <
\abs{\Re{\lambda_n}}$. From this definition it follows that
$\lambda_0=0$ and $\sss=\eig{0}/\Tr{\eig{0}}$. 
Equation~\eqref{Eq:eigenvectordecomposition} proves that in absence of LEPs the dynamics of
an open system can be understood in terms of a sum of complex-exponential decays 
towards the steady state.

Moreover, we recall
here some useful properties of the eigenmatrices
\cite{MingantPRA18_Spectral}:

\begin{lemma} \label{Lemma:Evolution_Eigenstate}
	Given Eq.~\eqref{eq:Liouvillian_Eigenstates}, $\exp{(\LL t)} \eig{i}=\exp{(\lambda_i t})\eig{i}$ .
\end{lemma}
Given Lemma~\ref{Lemma:Evolution_Eigenstate}, and since the Liouvillian is a trace-preserving map, it follows that
\begin{lemma} \label{Lemma:Eigenmatrices_are_traceless}
	If $\lambda_i \neq 0$, $\Tr{\eig{i}}=0$. 
	Moreover, if $\Tr{\eig{i}}\neq 0$, then $\lambda_i = 0$.
\end{lemma}
\begin{lemma}\label{Lemma:Real_is_Hermitian}
	If $\LL\eig{i}=\lambda_i\eig{i}$ then  $\LL\eig{i}^\dagger=\lambda_i^*\eig{i}^\dagger$.\\
	Thus, if $\eig{i}$ is Hermitian, then $\lambda_i$ has to be real.
	Conversely, if $\lambda_i$ is real and of degeneracy 1, $\eig{i}$ is Hermitian.
	If $\lambda_i$ has geometric multiplicity $n$, it is always possible to construct $n$ Hermitian eigenmatrices of $\LL$ with the eigenvalue $\lambda_i$ \footnote{\label{prova} The algebraic multiplicity of $\lambda$ is defined as the number of times $\lambda$ appears as a root of the characteristic equation. The geometric multiplicity, instead, is the maximum number of linearly independent eigenvectors associated with $\lambda$.}.
\end{lemma}

\subsection{Liouvillian Exceptional Points and time dynamics towards the steady state}
In this formalism, a Liouvillian exceptional point (LEP) is the point of the parameter space where two eigenmatrices 
of the Liouvillian coalesce. Since LEPs are associated with a nondiagonalizable Liouvillian, at the critical point $\LL$ can be written in its Jordan canonical form.
Consider now a Liouvillian admitting an EP of order 2. The equation $(\mathcal{L}-\lambda_{\rm EP}) \eig{{\rm EP}}=0$, $\lambda_{\rm EP}$ being the eigenvalue associated with the LEP, admits only
one nontrivial solution $\eig{{\rm EP}}^{(1)}$. There exists, however, a generalized eigenmatrix $\eig{{\rm EP}}^{(2)}$, which, by writing
the Liouvillian in its canonical form, is defined by $(\mathcal{L}-\lambda_{\rm EP})
\eig{{\rm EP}}^{(2)}=A \eig{{\rm EP}}^{(1)}$. We stress that the coefficient $A$ need not be equal to 1, since we choose $|\eig{{\rm EP}}^{(2)}|^2 =1$ \cite{Wiersig2016}.

The presence of the LEP has consequences on the dynamics towards the steady state. 
Indeed, at an LEP Eq.~\eqref{Eq:eigenvectordecomposition} becomes
\begin{equation}\label{rho_diag_EP}  
\hat\rho(t)=\sum\limits_i c_i(t)\hat\rho_i + c^{(1)}_{\rm
	EP}(t)\hat\rho^{(1)}_{\rm EP}+ c^{(2)}_{\rm EP}(t)\hat\rho^{(2)}_{\rm EP},
\end{equation}
where
\begin{equation}
c_{\rm EP}^{(1)}(t)= \exp\left(\lambda_{\rm EP}t\right){\rm
	Tr}[\hat\sigma^{(1)}_{\rm EP}\hat\rho(0)],
\end{equation}
while
\begin{equation}
c^{(2)}_{\rm EP}(t)= t \exp\left(\lambda_{\rm EP}t\right){\rm
	Tr}[\hat\sigma^{(2)}_{\rm EP}\hat\rho(0)].
\end{equation}
Contrary to Eq.~\eqref{Eq:eigenvectordecomposition},  Eq.~\eqref{rho_diag_EP} is not a sum of complex-exponential decays.
Thus, the presence of an exceptional point is captured by a polynomial times a decaying complex exponential. 
In this regard, the study of dynamical quantities can signal the occurrence of an LEP \cite{ArkhipovarXiv19,MingantPRA18_Spectral,SarandyPRA05_Adiabatic}.
While in this example we considered an LEP of order 2, presenting a behavior $t \exp\left(\lambda_{\rm EP}t\right)$, 
an LEP of order $n$ will present a $(n-1)$ degree polynomial dependence in the decay towards the steady state.

\subsection{Relation between the spectra of $\hat{H}_{\rm eff}$, $\LL'$, and $\LL$}

As we previously discussed, we have three possible mechanisms to
describe the dynamics of an open system: the NHH $\hat{H}_{\rm
	eff}$, the Liouvillian without quantum jumps $\LL'$, and the full
Liouvillian $\LL$. Since EPs are indicated by the spectra of these
three objects, a question arises about the relations between them.

Let us call $\ket{\phi_j}$ the right eigenvectors of $\hat{H}_{\rm
	eff}$, whose eigenvalues are $h_j$. We have
\begin{equation}\label{Eq:val_NHH}
\begin{split}
\hat{H}_{\rm eff}  \ket{\phi_j} &= h_j \ket{\phi_j}, \\
\bra{\phi_j} \hat{H}_{\rm eff}^\dagger  &= \left( \hat{H}_{\rm
	eff} \ket{\phi_j}  \right)^\dagger = h_j^* \bra{\phi_j}.
\end{split}
\end{equation}
First, let us investigate the relation between the spectral properties of $\LL'$ and of $\hat{H}_{\rm eff}$. Let us assume that ${\eig{j}'=
\ket{\phi_l}\bra{\phi_m}}$. From Eq.~\eqref{Eq:val_NHH}, it follows that
\begin{equation}
\LL'\eig{j}'=- i \left(\hat{H}_{\rm eff} \eig{j}' - \eig{j}' \hat{H}_{\rm eff}^\dagger  \right) =- i(h_l - h_m^*) \eig{j}'=\lambda_j' \eig{j}'.
\end{equation}
The set of eigenvectors of $\LL'$ is thus given by
$\eig{j}'=\{\ket{\phi_l}\bra{\phi_m}\}$. Thus, the onset of an HEP
is biunivocally determined by that of $\LL'$.
We stress that, to be consistent, one should compare $\Re{\lambda_i'}$ with  $\Im{h_i}$, due to the $(-i)$ factor in Eq.~\eqref{Eq:Linblad_semiclassical}.
The introduction of $\LL'$ allows a easier connection between $\hat{H}_{\rm eff}$ and $\LL$.

Now, we consider the spectra of $\LL'$ and $\LL$. Consider an eigenmatrix $\eig{j}'$ of $\mathcal{L}'$. We see
that
\begin{equation}
\begin{split}
\LL\eig{j}' & = - i \left(\hat{H}_{\rm eff} \eig{j}' - \eig{j}' \hat{H}_{\rm eff}^\dagger  \right) + \sum_\mu \hat{\Gamma}_\mu \eig{j}'\hat{\Gamma}_\mu' \\
&= - i(h_l - h_m^*) \eig{j}' + \sum_\mu \hat{\Gamma}_\mu \eig{j}'\hat{\Gamma}_\mu^\dagger.
\end{split}
\end{equation}
Therefore, all the eigenmatrices of $\LL$ are identical to those
of $\LL'$ if for each $\hat{\Gamma}_\mu$ it holds that $\hat{\Gamma}_\mu
\eig{j}'\hat{\Gamma}_\mu^\dagger \propto \eig{j}'$. In other
words, this condition is verified if $\hat{\Gamma}_\mu
\ket{\phi_l}\bra{\phi_m}\hat{\Gamma}_\mu^\dagger \propto
\ket{\phi_l}\bra{\phi_m}$. We conclude that if $\ket{\phi_l}$ and
$\ket{\phi_m}$ are right eigenvectors of each $\hat{\Gamma}_\mu$
for each $m$ and $l$, the spectrum of $\mathcal{L}'$ is
biunivocally determined by that of $\mathcal{L}$. The other
possibility is that we have two eigenmatrices $\eig{j}'$ and
$\eig{k}'$ such that $\LL' \eig{j}'=\lambda_j \LL' \eig{j}'$ and
$\LL' \eig{k}'=\lambda_j \LL' \eig{k}'$. In this case, the effect
produced by each quantum jump term $\mathcal{J}[\hat{\Gamma}_\mu]$ must be only to mix the eigenmatrices. But this
is equivalent to say that $\hat{\Gamma}_\mu$ and $\hat{H}_{\rm
	eff}$ share an eigenvector basis, and, therefore, can be
simultaneously diagonalized. Therefore we have the following Lemma:
\begin{lemma}\label{Lemma:When_NHH_is_true}
	If $[\hat{\Gamma}_\mu, \hat{H}_{\rm eff}]=0$, the eigenmatrices $\eig{j}$ of $\LL$ are of the form $\ket{\phi_l}\bra{\phi_m}$, where $\ket{\phi_l}$ is an eigenvector of $\hat{H}_{\rm eff}$.
	The eigenvalues of $\LL$ are $\lambda_i = - i(h_l - h_m^*) + g_m g_l^*$, where $\hat{\Gamma}_\mu \ket{\phi_m}=g_m \ket{\phi_m}$.
\end{lemma}
From now on, with a slight abuse of notation, we will say that
$\hat{H}_{\rm eff}$ has the same eigenvectors of $\LL$ if
$\eig{j}=\ket{\phi_l}\bra{\phi_m}$.

As an example of Lemma~\ref{Lemma:When_NHH_is_true}, let us consider the bosonic
annihilation operator $\hat{a}$, and the following Liouvillian:
\begin{equation}
\left\lbrace
\begin{split}
\LL &= - i [\hat{H}, \bigcdot] + \frac{\gamma}{2} \mathcal{D}[\hat{a}^\dagger \hat{a}], \\
\hat{H} &= \omega \hat{a}^\dagger \hat{a}.
\end{split}    \right.
\end{equation}
Clearly, $[\hat{H}_{\rm eff}, \hat{a}^\dagger \hat{a}] = 0$, where
$\hat{H}_{\rm eff}= \omega \hat{a}^\dagger \hat{a} - i
\frac{\gamma}{2} \hat{a}^\dagger \hat{a}$. The eigenvalues of
$\hat{H}_{\rm eff}$ are the number (Fock) states
$\ket{n}$, and its eigenvalues are $h_n = (\omega - i
\frac{\gamma}{2} n) n$. We conclude that the Liouvillian eigenstates
are $\eig{j}=\ket{m}\bra{n}$, whose eigenenergies are $- i
\omega(m - n) - \frac{\gamma}{2} (m-n)^2 $.

\subsection{Go and no-go theorems for the equivalence of LEPs and HEPs}
\label{sec:nogo} The question about EPs in the
quantum case can now be partially addressed. 
Indeed, as stated in  Lemma~\ref{Lemma:When_NHH_is_true}, there are cases in which the eigenvectors of the Liouvillian are identical to that of the NHH.
Given this property, is it possible to observe an HEP whose generalized eigenvectors are identical to those of $\LL$?

\subsubsection{A no-go theorem in the quantum regime}
As we stressed in Sec.~\ref{Sec:Spectrum}, the steady-state density matrix plays a central role in the description of open quantum systems.
Therefore, one may wonder whether it is possible to observe some LEPs associated with the steady state $\sss$. In
Refs.~\cite{MingantPRA18_Spectral,AlbertPRA14,BaumgartnerJPA08}
the following Lemma was demonstrated:

\begin{lemma}\label{Lemma:No_Jourdan_in_Zero}
	If $\lambda_i=0$ has degeneracy $n$, then there exists $n$ independent right eigenvectors and $n$ independent left eigenvectors of the Liouvillian (the algebraic multiplicity of $\lambda_i$ is identical to the geometrical one).
\end{lemma}

This proposition has profound consequences on the structure of a
Liouvillian spectrum and on the presence of EPs. Indeed, since EPs
require a Jordan canonical form, there cannot be any EPs for those
eigenmatrices of a Liouvillian whose eigenvalue is zero.
Therefore, an EP can exist in the ``excited'' states of a
Liouvillian (that is, $\eig{i}$ whose $\Re{\lambda_i}<0$ and which represent the dynamical decay of an initial state towards its steady state). 
For a quantum two-level system (e.g., a spin $\frac{1}{2}$), this means that no NHH
exhibiting an EP correctly captures the underlying physics, and
quantum jumps must necessarily be taken into account. Indeed, the
NHH would have two eigenvectors $\ket{\phi_{1}}$ and $\ket{\phi_{2}}$, which coalesce
at the HEP. 
However, if the spectrum of $\mathcal{L}$ was to coincide with that of
$\mathcal{L}'$, and $\LL$ admit a steady state $\sss$,
we conclude that $\sss$ would be the exceptional point, proving the contradictory affirmation that there would be a LEP in the steady state (examples in Secs.~\ref{Sec:Example_1}~and \ref{Sec:Example_2}).

The previous result can be generalized to systems with more than
two levels (but still, of a finite dimension). 
Let us assume now that there exists an HEP of order 2 (the demonstration is similar for higher-order EPs).
Therefore, we have $\hat{H}_{\rm eff} \ket{\phi_1}=h_1 \ket{\phi_1}$ and
$\hat{H}_{\rm eff} \ket{\phi_2}=h_1 \ket{\phi_1} + A \ket{\phi_2}$
(where the factor $A$ ensures the correct normalization of $\ket{\phi_2}$ \cite{Wiersig2016}).
The proof can be outlined as follows: \begin{itemize}
	\item  In the subspace spanned by $\ket{\phi_1}$ and $\ket{\phi_2}$,
	$\mathcal{L}'$ is a $4\times4$ matrix presenting an LEP', and there is a true eigenvector $\eig{1}=\ket{\phi_1} \bra{\phi_1}$, such that $\mathcal{L}'\eig{1}=-i (h_1-h_1^*) \eig{1}$.
	\item We proceed by contradiction (reductio ad absurdum), and we assume that $\eig{1}$ is an eigenmatrix of $\LL$ such that $\LL\eig{1}=\lambda_{1}\eig{1}$.
	Moreover, that $\lambda_{1}$ is associated with a LEP.
	\item From Lemma~\ref{Lemma:No_Jourdan_in_Zero}, since there is a LEP, we deduce that $\lambda_{1}\neq 0$.
	\item Since $\eig{1}=\ket{\phi_1} \bra{\phi_1}$, we have that $\Tr{\eig{1}}=1$.
	From Lemma~\ref{Lemma:Eigenmatrices_are_traceless}, we conclude that $\lambda_{1}=0$.
\end{itemize}
Thus, we immediately arrive at a contradiction.
This demonstration can be easily generalized to higher-order EPs and to degeneracies in the NHH spectrum.
We conclude the following:
\begin{theorem}\label{Thm:no_HEP_are_LEP}
	In the quantum limit, a given NHH exhibiting exceptional points
	cannot have exactly the same spectral structure (i.e., eigenvalues and eigenmatrices) as the corresponding full Liouvillian.
\end{theorem}

We notice that Theorem~\ref{Thm:no_HEP_are_LEP} is valid only for $\LL$ and not for $\LL'$.
Indeed, for $\LL'$ it is possible to have $\Re{\lambda_{i}}>0$, and $\LL'$ is not a trace-preserving superoperator. Therefore, Lemma~\ref{Lemma:Eigenmatrices_are_traceless} does not hold for $\LL'$.
Moreover, we remark that LEPs and HEPs may become partially equivalent in the semiclassical regime, as we will discuss in the following section.

Finally, there is another intriguing consequence of Theorem 1. Indeed, in the quantum
regime, one can have LEPs without any Hamiltonian counterparts. In this case, the effect of $\mathcal{J}$ is
not detrimental but necessary to produce EPs.
Indeed, one can have an exceptional point describing the decay of the system coherences, which cannot be captured by an NHH approach (an actual example is discussed in Sec.~\ref{Sec:Example_1}).

\subsubsection{Equivalence of LEPs and HEPs in the semiclassical limit}
\label{sec:when_NHH_is_L}

Let us summarize the three main results we have obtained in the previous sections:
\begin{enumerate}[label=(\roman*)]
	\item Lemma \ref{Lemma:When_NHH_is_true}: $\LL$ has the same eigenvectors of
	$\hat{H}_{\rm eff}$ if every jump operator commutes with the
	effective Hamiltonian.
	\item Lemma~\ref{Lemma:No_Jourdan_in_Zero}: Any effect of an EP has a dynamical nature and, in a
	time-independent Lindblad master equation or in a time-independent
	NHH, some effects can be observed only in the transient dynamics of the
	system towards its steady state. We stress that this does not mean
	that the system is undriven or in the ground state of its
	NHH. No matter the details of the processes present in the
	Hamiltonian and the dissipators, at the Lindblad master-equation
	level the effect of EPs cannot be observed in the steady state.
	\item Theorem~\ref{Thm:no_HEP_are_LEP}: The structure of the spectrum of a
	Liouvillian EP (LEP) cannot be identical to that of a Hamiltonian EP (HEP).
\end{enumerate}

Do these observations (i)--(iii) mean that there cannot be any correspondence
between the Liouvillian and NHH eigenvectors? The answer is no,
but only if we consider that the effect of quantum jumps, in some
particular subspace, can be such as to combine only certain wave functions obtained via the NHH approach. In this regard, the spectral properties are not exactly identical, but the physics described by the NHH can capture the same phenomena of the Liouvillian. There is, however, an additional caveat regarding the way in which a semiclassical limit should be approached.
Indeed, the term $\mathcal{J}\rhot$ in
Eq.~\eqref{Eq:Evolution_non_hermitian} is often omitted under the
assumption that the \emph{steady state} is a semiclassical state.
However, nothing guarantees that \emph{all the eigenmatrices}
$\eig{j}$ of $\LL$ are all compositions of semiclassical states.
Indeed, we should verify whether the effect of quantum jumps on the
set of matrices $\eig{i}'$ is negligible, $\eig{i}'$ being the
eigenstates of $\LL'$. It may happen that the overall effect of
quantum jumps is either to mix the state trivially, thus
retrieving similar features, or to mix them in a nontrivial way,
producing different effects.

We can formalize this intuition in a more rigorous way. Since we are interested in capturing only the behavior of a certain ``semiclassical'' part of
the spectrum, let us assume that we want to know if a set of
eigenvectors correctly approximates part of the Liouvillian dynamics. In
this case, we construct the set eigenmatrices of $\LL'$, i.e.,
\begin{equation}
\eig{i=l+m}'=\ket{\phi_{l}}\bra{\phi_{m}},
\end{equation}
and we have
\begin{equation}
\LL \eig{i}' = \lambda'_i \eig{i}' + \mathcal{J}[\hat{\Gamma}_\mu] \eig{i}'.
\end{equation}
We write the effect of the quantum jumps $ \mathcal{J}[\hat{\Gamma}_\mu]$ as a component along
$\eig{i}'$ and a residue $\hat{\sigma}_i$, so that ${\mathcal{J}[\hat{\Gamma}_\mu] \eig{i}' = j_i
\eig{i}' + \hat{\sigma}_i}$. In the limit in which ${\left\Vert
\hat{\sigma}_i \right\Vert \ll \left\Vert (\lambda'_i + j_i')
\eig{i}' \right\Vert}$ (or $ \hat{\sigma}_i$ is exactly zero), we
can approximate the short-time dynamics of the full Liouvillian
with the effective Hamiltonian for any superposition of matrices
$\eig{i}$. Therefore, we are requiring that
\begin{equation}\label{Eq:Semiclassical}
\hat{\Gamma}\ket{\phi_l} \bra{\phi_m} \hat{\Gamma}^\dagger =
\left(g_l \ket{\phi_l} + \epsilon_l \ket{\varphi_l}\right) \left(g_m
\bra{\phi_m} + \epsilon_m \bra{\varphi_m}\right),
\end{equation}
where $\ket{\varphi_l}$ is the action outside of the eigenvector
space. 
In case $|g_l \epsilon_m|\ll1$ and $|g_m \epsilon_l|\ll1$, the semiclassical condition is satisfied and the dynamics is described by an NHH.
There is, however, a case in which Eq.~\eqref{Eq:Semiclassical}
becomes exact. Specifically, if there exist $\ket{\phi_0}$ such that $\hat{\Gamma}\ket{\phi_0}=0$, Eq.~\eqref{Eq:Semiclassical} is true for any $\bra{\phi_m}$. Therefore, we have
the following:
\begin{theorem}
	\label{Thorem:When_NHH_is_true_for_EPs}
	Let $\ket{\phi_{0}}$ be a eigenvector of the NHH $\hat{H}_{\rm eff}$ such that $\hat{\Gamma}\ket{\phi_{0}}=0$.
	In this case, the Liouvillian has a set of eigenmatrices $\ket{\phi_{0}}\bra{\phi_{m}}$ and $\ket{\phi_{m}}\bra{\phi_{0}}$, where $\hat{H}_{\rm eff} \ket{\phi_{m}} = h_m \ket{\phi_{m}}$.
\end{theorem}
One may be surprised by the fact that
$\ket{\phi_{0}}\bra{\phi_{m}}$ is, somehow, a
semiclassical limit. 
However, $\ket{\phi_{0}}$ represents the
vacuum of the jump operator, and therefore
$\ket{\phi_{0}}\bra{\phi_{m}}$ describes the continuous decay of the state towards the vacuum, due to the environment absorbing its energy. 
If we consider now a semiclassical state $\bra{\phi_{m}}$, according to the semiclassical theory of an NHH,
its norm must decay until it becomes zero. This is exactly what is
predicted by $\ket{\phi_{0}}\bra{\phi_{m}}$. Indeed, having
generalized the NHH to the Liouvillian context by adding the
quantum jumps terms makes impossible for an actual density matrix $\ket{\phi_{m}}\bra{\phi_{m}}$ to lose
its norm. In this regard, we expect that a model, which satisfies
Theorem~\ref{Thorem:When_NHH_is_true_for_EPs} is a semiclassical
model.

We stress that, again, the conditions of
Eq.~\eqref{Eq:Semiclassical} cannot exactly be satisfied for any
arbitrary pair of matrices $\ket{\phi_l} \bra{\phi_m}$,
otherwise it would imply the existence of multiple steady states
with an EP, disproving Lemma~\ref{Lemma:No_Jourdan_in_Zero} and Theorem~\ref{Thm:no_HEP_are_LEP}.

We also notice that Theorem~\ref{Thorem:When_NHH_is_true_for_EPs} is a sufficient (but not necessary) condition to have the correspondence between 
the eigenvalues of $\LL$ and those of NHH.
There may exist NHHs whose eigenvalues and eigenvectors are equal to a subset of those of the corresponding Liouvillians, even without obeying Theorem~\ref{Thorem:When_NHH_is_true_for_EPs}.

Finally, there exists NHHs and Liouvillians which display HEPs and LEPs for the same combination of parameters (e.g., in \cite{ArkhipovarXiv19}).
This does not imply that the eigenvectors stemming from the two procedures must be identical.

\subsection{Physical meaning of the Liouvillian eigenmatrices}
\label{Sec:Spectral_decomposition}

To address the correspondence between LEPs and HEPs, one has
to correctly interpret the physical meaning of the Liouvillian
eigenmatrices. Here, we provide a pedagogical discussion, following
that of Refs.~\cite{MingantPRA18_Spectral,RivasBOOK_Open}.

\subsubsection{The case of a real Liouvillian eigenvalue $\lambda_i$}

When $\lambda_i$ is real, $\eig{i}$ can be constructed to be
Hermitian (see Lemma~\ref{Lemma:Real_is_Hermitian}). By
diagonalizing it, one obtains the spectral decomposition
\cite{RivasBOOK_Open, MingantPRA18_Spectral}:
\begin{equation}
\eig{i}= \sum_n p_n^{(i)} \ket{\psi_n^{(i)}}\bra{\psi_n^{(i)}},
\end{equation}
where $\braket{\psi_n^{(i)}|\psi_m^{(i)}}=\delta_{n,m}$. Since all
the coefficients $p_n^{(i)}$ must be real and since $\eig{i}$ is
traceless (see
Lemmas~\ref{Lemma:Eigenmatrices_are_traceless}~and~\ref{Lemma:No_Jourdan_in_Zero}),
we can sort $p_n$ in such a way to have $p_n^{(i)}>0$ for
$n\leq\bar{n}$, and $p_n^{(i)}<0$ for $n> \bar{n}$. Thus, we have
\begin{equation}\label{Eq:DecoRhoReal}
\hat{\rho}_i \propto \hat{\rho}_i^+ - \hat{\rho}_i^-,
\end{equation}
where
\begin{eqnarray}
\hat{\rho}_i^+&=&\sum_{n\leq\bar{n}} p_n^{(i)}
\ket{\psi_n^{(i)}}\bra{\psi_n^{(i)}} , \cr &&\cr \hat{\rho}_i^-
&=& - \sum_{n>\bar{n}} p_n^{(i)}
\ket{\psi_n^{(i)}}\bra{\psi_n^{(i)}},
\end{eqnarray}
and the coefficients $\{p_n\}$ have been normalized to ensure
$\Tr{\hat{\rho}_i^+}=\Tr{\hat{\rho}_i^-}=1$. With this definition,
$\eig{i}^\pm$ are density matrices. \emph{The wave functions associated to} $\eig{i}^\pm$ \emph{are those that can be compared to
	the $\ket{\phi_i}$ characterizing an NHH.}

\subsubsection{The case of a complex Liouvillian eigenvalue $\lambda_i$}
Let us now consider a right eigenmatrix $\eig{i}$ with a complex
eigenvalue $\lambda_i$. As it stems from
Eq.~\eqref{Eq:eigenvectordecomposition}, to ensure that
$\hat{\rho}(t)$ is an Hermitian eigenmatrix, $\eig{i}$ must always
appear in combination with its Hermitian conjugate
$\eig{i}^\dagger$, which is also an eigenmatrix of $\LL$
(Lemma~\ref{Lemma:Real_is_Hermitian}). Thus, one can simply
consider the Hermitian combinations: symmetric $\eig{i}^{\rm
	s}=\eig{i}+\eig{i}^\dagger$ and antisymmetric $\eig{i}^{\rm a}=i
\, (\eig{i}-\eig{i}^\dagger)$. By performing again an
eigendecomposition of those states, we obtain $\eig{i}^{\rm s}=
\eig{i}^{\rm s\, +} - \eig{i}^{\rm s\, -}$ and $\eig{i}^{\rm a}=
\eig{i}^{\rm a\, +} - \eig{i}^{\rm a\, -}$.

\section{Example of Theorem 1: a system with LEPs but without HEPs}
\label{Sec:Example_1}
\begin{figure}
	\centering
	\includegraphics[width=.99\linewidth]{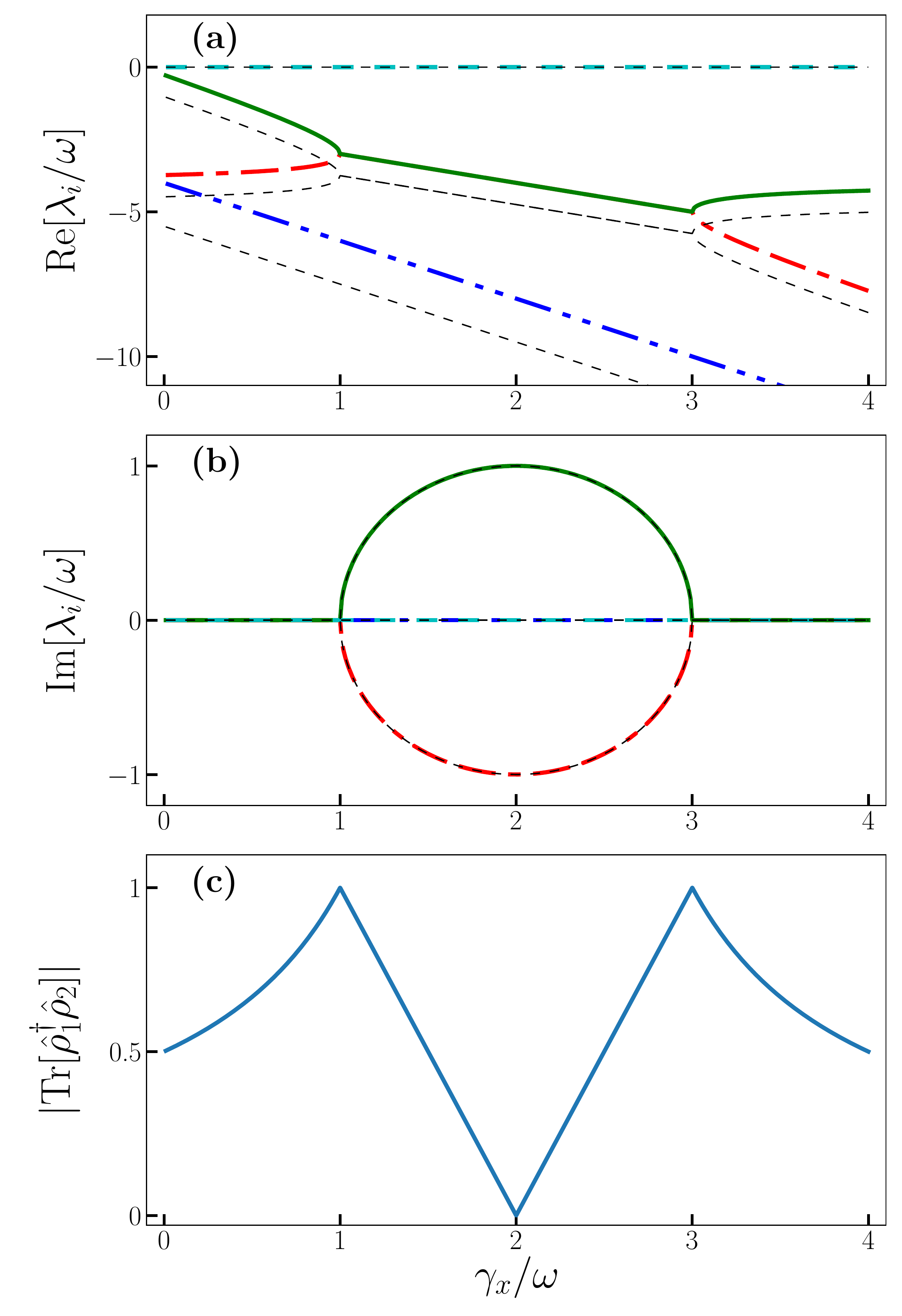}
	\caption{Spectral properties of the Liouvillian in Eq.~\eqref{Eq:Spin_liouvillian} in the case of $\gamma_y=2\omega$. This Liouvillian has a LEP, but the associated NHH has no HEP. 
		The thick curves represent $\gamma_-=0$, while the thin dashed ones are for $\gamma_-=\omega$.
		(a)~Real and (b)~imaginary parts of the Liouvillian eigenvalues in Eq.~\eqref{Eq:Spin_two_axes_vals} as a function $\gamma_x/\omega$, i.e., the dissipation in the $\hat{\sigma}_x$ direction rescaled by the qubit energy $\omega$.
		(c)~Scalar product between the two eigenmatrices $\eig{1}$ and $\eig{2}$ in Eq.~\eqref{Eq:Spin_two_axes_eigen} as a function of $\gamma_x/\omega$.}
	\label{fig:exceptionalspin2axes}
\end{figure}

In this section, we address the question of whether there exists
any model exhibiting LEPs but not HEPs, thus confirming Theorem~\ref{Thm:no_HEP_are_LEP}.
At first, let us consider a rather general model of a spin-$1/2$,
with Hamiltonian
\begin{equation}
\hat{H} = \frac{\omega}{2} \hat{\sigma}_z,
\end{equation}
which evolves under the action of the three competing decay
channels, ($\hat\sigma_x$, $\hat\sigma_y$, and $\hat\sigma_-$)
described by
\begin{equation}\label{Eq:Spin_liouvillian}
\LL \rhot= - i [\hat{H}, \rhot] + \frac{\gamma_- }{2}
\DD[\hat{\sigma}_-] \rhot
+\frac{\gamma_x}{2}\mathcal{D}[\hat\sigma_x]
+\frac{\gamma_y}{2}\mathcal{D}[\hat\sigma_y],
\end{equation}
$\hat\sigma_{x,\, y, \, z}$ being the Pauli matrices, $\hat{\sigma}_\pm = (\sigma_{x} \pm i \sigma_{y})/2$ and
$\hat{\sigma}_z=\hat{I}-2\hat{\sigma}_+\hat{\sigma}_-$.
Since this master equation is invariant under the exchange
$\hat{\sigma}_- \to -\hat{\sigma}_-$, this model explicitly
presents a $\mathcal{Z}_2$ symmetry
\cite{AlbertPRA14,MingantPRA18_Spectral}. Moreover, there are
several terms which can compete in determining the relaxation rate
towards the steady state: (i) the Hamiltonian oscillations; (ii)
the dissipation along the $x$ and $y$ axes; and (iii) the spin flips
described by $\hat{\sigma}_-$.

First, we note that the NHH structure is trivial, since
$\hat{H}_{\rm eff}$ is already diagonal in the $\hat{\sigma}_z$
basis, and its matrix form reads
\begin{equation}\label{Eq:Semiclassical_NHH_wrong}
\hat{H}_{\rm eff}=\frac{1}{2}\left(
\begin{array}{cc}
\omega -i \gamma_{x}- i \gamma_{y} -i \gamma_- & 0 \\
0 & -\omega -i \gamma_{x} - \gamma_{y}\\
\end{array}
\right). ,
\end{equation}
This equation cannot present any EP, since no change in parameters
can make the two eigenvalues equal.

Nevertheless, the Liouvillian can present several interesting
properties. We have
\begin{equation}\label{Eq:Spin_two_axes_vals}
\left\{\begin{split}
\lambda_0 &= 0,\\
\lambda_{1, \, 2} &= -\frac{\gamma_-}{2}-\gamma _x -\gamma_y\pm\Omega, \\
\lambda_3&=\gamma _--2 \left(\gamma_y+\gamma_x\right),
\end{split}\right.
\end{equation}
and
\begin{equation}\label{Eq:Spin_two_axes_eigen}
\left\{\begin{split} \eig{0}&\propto\rho^{SS}=\frac{1}{2 \gamma
	_x+ 2 \gamma _y +\gamma _-}\begin{pmatrix}
\gamma _x+\gamma _y & 0 \\
0 & \gamma _x+\gamma _y +\gamma _-
\end{pmatrix},  \\
\eig{1, \, 2}&\propto \begin{pmatrix} 0 & -i
\omega \pm \Omega \\
\gamma _x-\gamma _y & 0 \\
\end{pmatrix}, \\
\eig{3}&\propto\begin{pmatrix}
-1 & 0 \\
0 & 1 \\
\end{pmatrix},
\end{split}\right.
\end{equation}
where $\Omega= \sqrt{\gamma_x^2+\gamma_y^2-2 \gamma_x \gamma_y
	-\omega ^2}$.

Therefore, in the case $\gamma_{y}>\omega$, this Liouvillian
exhibits two EPs, one for $\gamma_{x}=\gamma_{y}-\omega$, and one
for $\gamma_{x}=\gamma_{y}+\omega$. We study this configuration
for $\gamma_-=0$ in Fig~\ref{fig:exceptionalspin2axes}. Here, the
key parameter is $\Gamma$ in $\lambda_{1, \, 2}$ and $\eig{1, \,
	2}$. Therefore, we can identify three regimes [cf.
Figs.~\ref{fig:exceptionalspin2axes}(a) and \ref{fig:exceptionalspin2axes}(b)]:
\begin{enumerate}[label=(\roman*)]
	\item The case of
	$\gamma_x<\gamma_{y}-\omega$, where the dynamics is dominated by
	the dissipation channel $\DD[\hat{\sigma}_y]$, and the decay
	towards the steady state is purely exponential.
	\item The case of
	$\gamma_{y}-\omega<\gamma_x<\gamma_{y}+\omega$, where the
	competition between the dissipation along the $\hat{\sigma}_x$ and
	$\hat{\sigma}_x$ directions allows for Hamiltonian oscillations
	towards the steady state.
	\item The case of
	$\gamma_x>\gamma_{y}+\omega$, when the dissipative dynamics is
	dominated by the damping in the $\hat{\sigma}_x$ direction.
\end{enumerate} 
This change in the spectral properties of the Liouvillian is
signaled by a coalescence of the eigenvectors, as shown in
Fig.~\ref{fig:exceptionalspin2axes}(c). The case of $\gamma_-\neq
0$ is also plotted in Fig.~\ref{fig:exceptionalspin2axes} with
dotted curves and shows similar spectral features, with a
remarkable difference: an overall shift of
$\gamma_-/2$ in $\lambda_{1, \, 2}$, and of $\gamma_-$ in
$\lambda_3$.

\subsubsection{Purely quantum exceptional points}

The study of the Liouvillian in Eq.~\eqref{Eq:Spin_liouvillian}
naturally raises the question of the meaning of the Liouvillian EPs
which are induced by quantum jumps and thus are not observed in
the corresponding NHH dynamics. Indeed, the term determining the
EP is $\hat{\sigma}_{x} \rhot \hat{\sigma}_{x}$. According to 
measurement theory, this process can be interpreted as the
backaction of a measurement apparatus on a system
\cite{Barnett_BOOK_Info}. In the present case, such ``reading
apparatus'' is the environment itself
\cite{Haroche_BOOK_Quantum,Wiseman_BOOK_Quantum}, which projects
the system on the eigenspace of its pointer states. In this
regard, this exceptional point is induced by a purely quantum
effect, and is really due to the measurement and not to the
``semiclassical'' dissipation caused by the environment. That is, the
bare presence of the measurement apparatus and the reading of it
induces quantum jumps.
Finally, $\eig{1,\, 2}$ represent the loss of coherence of the density matrix into the environment.
As such, these LEPs are due to this purely quantum effect, and cannot be explained
by the semiclassical approximation in Eq.~\eqref{Eq:Semiclassical_NHH_wrong}.

\begin{figure}
	\centering
	\includegraphics[width=.99\linewidth]{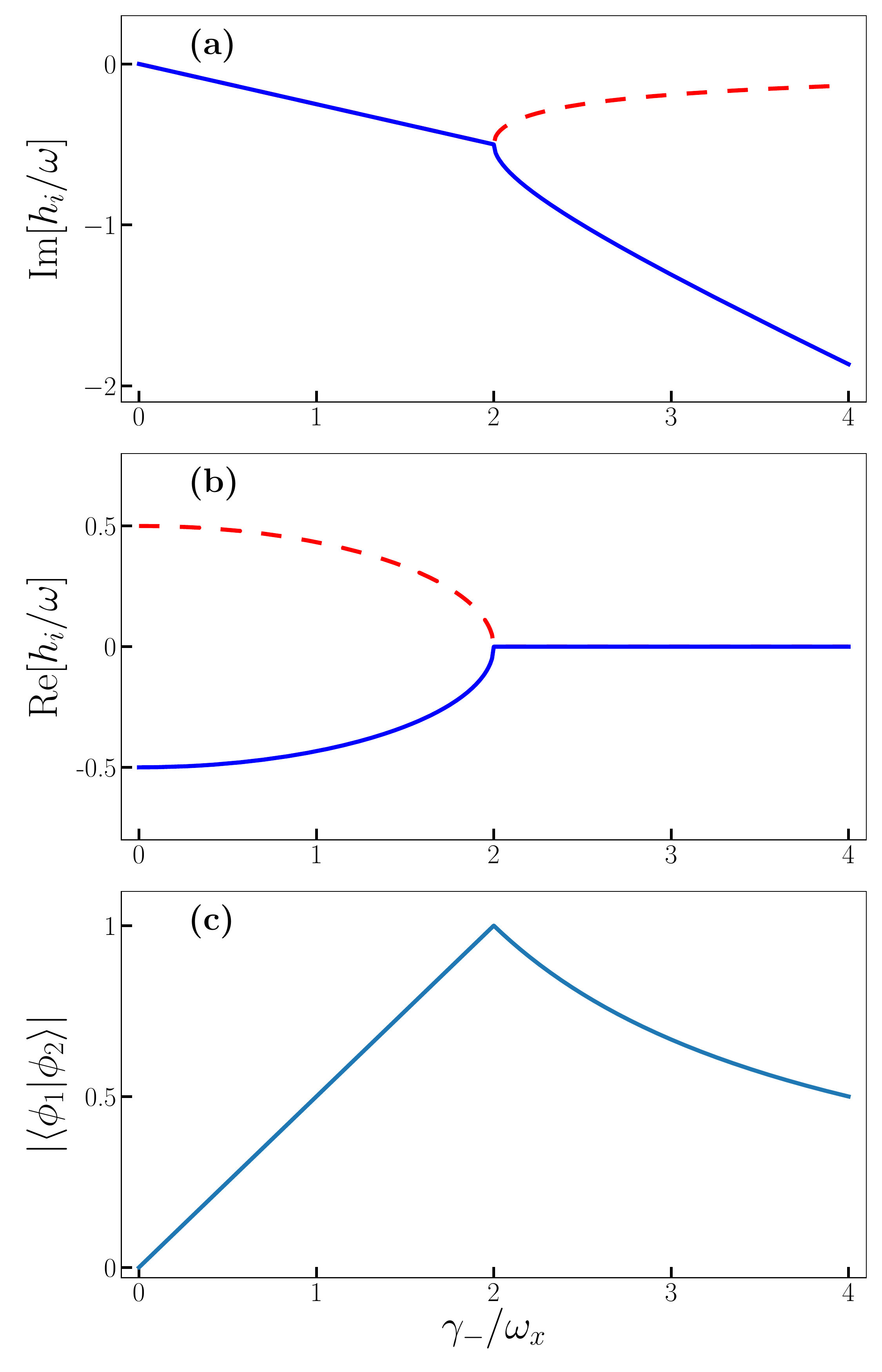}
	\caption{Spectral properties of the NHH in Eq.~\eqref{Eq:NHH_with_EP} showing an HEP.
		(a) Imaginary and (b) real parts of the eigenvalues in Eq.~\eqref{Eq:spectrum_NHH_spin_with_HEP} as a function of $\gamma_-/\omega_x$: the ratio between the spin-flip rate and the drive.
		(c) Scalar product between $\ket{\phi_1}$ and $\ket{\phi_2}$, given in Eq.~\eqref{Eq:Ham_eigen_EP}, as a function of $\gamma_-/\omega_x$.}
	\label{fig:driven_Hamiltonian}
\end{figure}

\begin{figure}
	\centering
	\includegraphics[width=.99\linewidth]{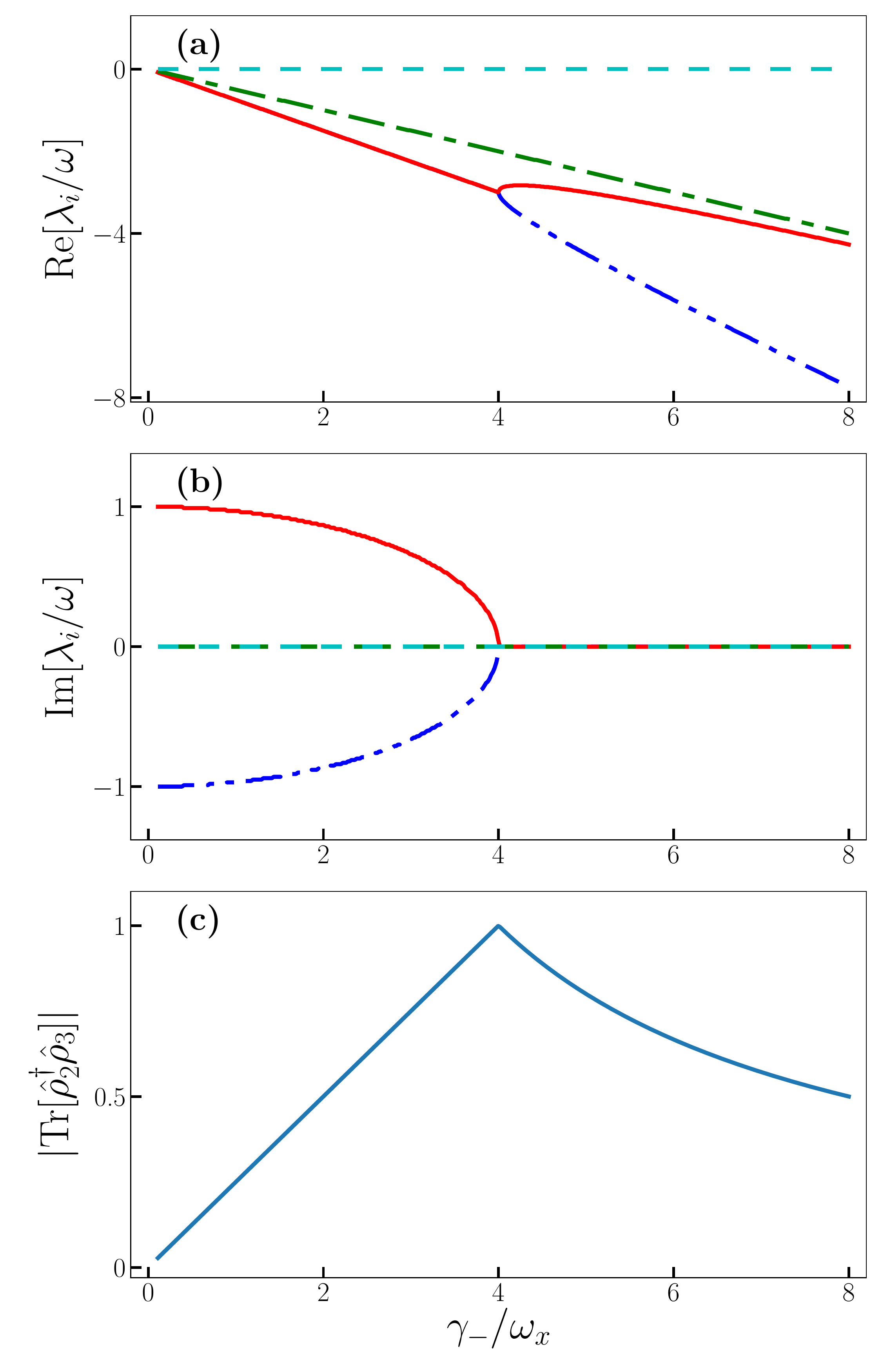}
	\caption{Spectral properties of the Liouvillian in Eq.~\eqref{Eq:Spin_liouvillian_with_NNHEP}, exhibiting a LEP, but for the same parameters as the NHH studied in Fig.~\ref{fig:driven_Hamiltonian}.
		(a) Real part and (b) imaginary parts of the Liouvillian eigenvalues in Eq.~\eqref{Eq:Liouvillian_spectrum_spin_system_with_HEP} as a function of $\gamma_-/\omega_x$.
		Note that according to Eq.~\eqref{Eq:Linblad_semiclassical}, one should compare $\Re{\lambda_i}$ with $\Im{h_i}$ in Fig.~\ref{fig:driven_Hamiltonian}(a), and $\Im{\lambda_i}$ with $\Re{h_i}$ in Fig.~\ref{fig:driven_Hamiltonian}(b).
		(c) Scalar product between the two eigenmatrices $\eig{2}$ and $\eig{3}$, given in Eq.~\eqref{Eq:Eigenmatrices_Liouvillian_driven}, as a function of $\gamma_-/\omega_x$.}
	\label{fig:driven_Liouvillian}
\end{figure}

\section{Example 2 of Theorem 1: a system with nonequivalent LEPs and HEPs}
\label{Sec:Example_2} Here we study a model of a single dissipative driven
spin exhibiting both LEPs and HEPs. This is another example
obeying the condition of Theorem~\ref{Thm:no_HEP_are_LEP}. This time, both the NHH and $\LL$ present EPs. 
However, by comparing them we find several discrepancies. 

This
model is described by
\begin{equation}
\hat{H} = \frac{\omega_x}{2} \hat{\sigma}_x,
\end{equation}
which evolves under the action of the following Liouvillian
decaying channel
\begin{equation}\label{Eq:Spin_liouvillian_with_NNHEP}
\LL \rhot= - i [\hat{H}, \rhot] + \frac{\gamma_- }{2}
\DD[\hat{\sigma}_-] \rhot.
\end{equation}
We stress that an experimental study of this system using postselection techniques has been presented in  Ref.~\cite{NaghilooNatPhys19}.

\subsection{The non-Hermitian Hamiltonian spectrum}

We begin our study by considering the following NHH
\begin{equation}\label{Eq:NHH_with_EP}
\hat{H}_{\rm eff} = \frac{\omega_x}{2} \hat{\sigma}_x - i
\frac{\gamma_-}{2} \hat{\sigma}_+ \hat{\sigma}_-,
\end{equation}
which results from Eq.~\eqref{Eq:Spin_liouvillian_with_NNHEP} if
we ignore the quantum jump term in $\DD[\hat{\sigma}_-]$. We
remark that, by the addition of a constant term $(-i \gamma_-
\mathds{1})$, this model becomes the celebrated two-level system
showing a parity-time ($\mathcal{PT}$)-symmetry breaking, extensively discussed
in, e.g., Refs. \cite{El-GanainyNature2018,Miri2019}.
Indeed, this Hamiltonian has eigenvalues
\begin{equation}\label{Eq:spectrum_NHH_spin_with_HEP}
h_{1, \, 2} =\frac{1}{4} \left(-i \gamma_{-} \mp \zeta \right),
\end{equation}
and eigenvectors
\begin{equation}\label{Eq:Ham_eigen_EP}
\ket{\phi_{1, \, 2}} \propto \left[i \gamma_- \mp \zeta \,, \quad
2 \omega_x\right],
\end{equation}
where $\zeta=\sqrt{4 \omega _x^2-\gamma_{-}^2}$.
The imaginary and real parts of the eigenvalues $h_i$ are plotted in Figs.~\ref{fig:driven_Hamiltonian}(a) and Figs.~\ref{fig:driven_Hamiltonian}(b), respectively. Notice that panel (a) represents the imaginary part of $h_i$, while panel (b) focuses on the real part.
In this way, one can directly compare the results for $\Im{h_i}$ with those $\Re{\lambda_i}$ in Figs.~\ref{fig:driven_Hamiltonian}(a)~and~\ref{fig:driven_Liouvillian}(a), taking into account the $(-i)$ factor in Eq.~\eqref{Eq:Linblad_semiclassical}.

For $\gamma_{-} / \omega_x = 2$, the eigenvalues of $\hat{H}_{\rm
	eff}$ are degenerate, $h_1=h_2$, and the corresponding
eigenvectors $\ket{\phi_{1}}$ and $\ket{\phi_{2}}$ of $\hat{H}_{\rm eff}$ coalesce
(see also Fig.~\ref{fig:driven_Hamiltonian}). Since this model
exhibits some ``semiclassical'' EPs, the natural question arises,
what happens to these EPs once the quantum jump term is taken into
account? Since $[\hat{\sigma}_-,\hat{\sigma}_x]\neq 0$ and $\hat{\sigma}_-\ket{\phi_{1, \, 2}}\neq 0$, we expect that
the spectral structure of these two models can be remarkably different (cf.
Lemma~\ref{Lemma:When_NHH_is_true} and Theorem~\ref{Thm:no_HEP_are_LEP}).

\begin{figure}
	\centering
	\includegraphics[width=.99\linewidth]{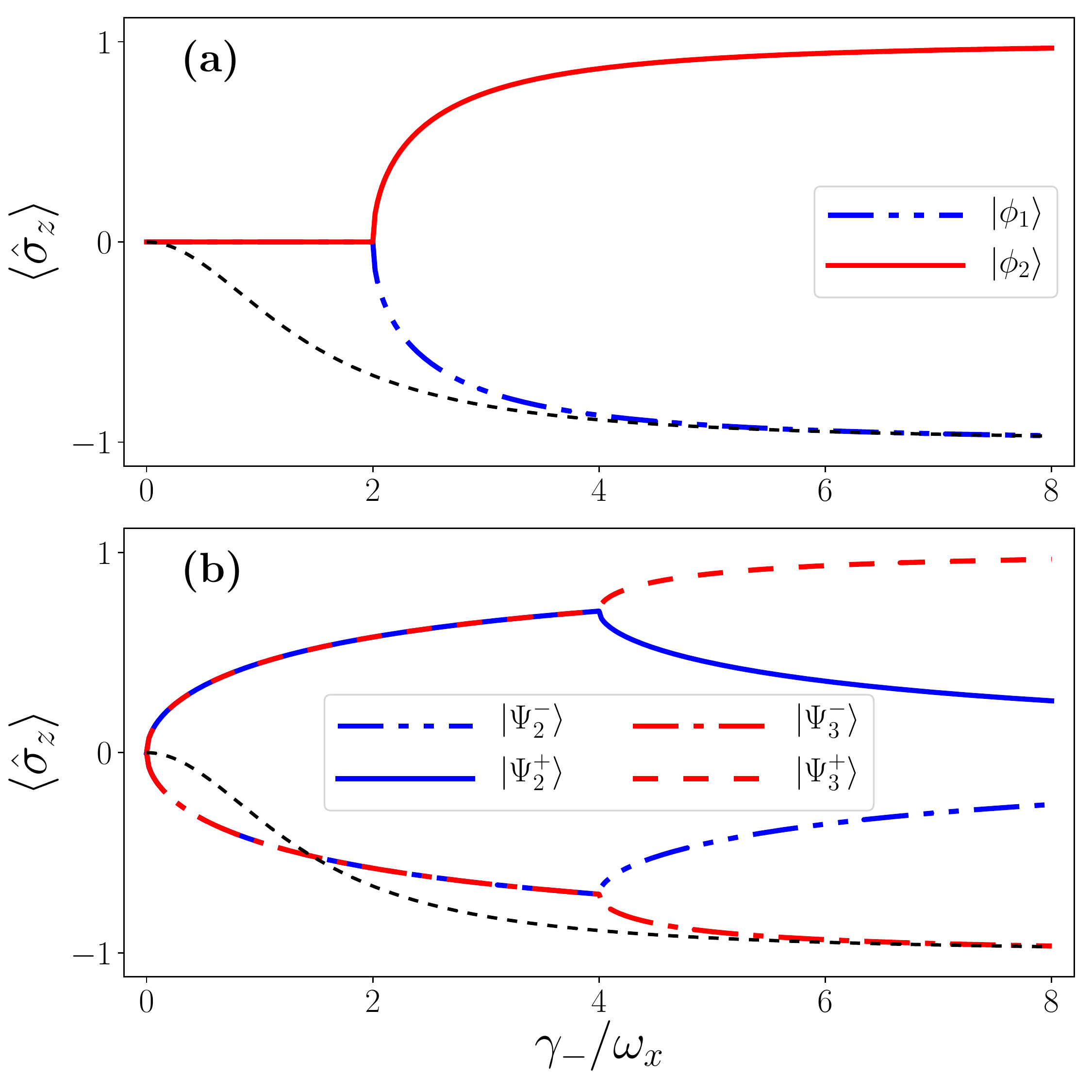}
	\caption{Expectation value of $\braket{\hat{\sigma}_z}$ as a function of $\gamma_{-}/\omega_x$. In both panels the dashed black curves represent the mean value of photons in the steady state.
		(a) NHH eigenvectors [Eq.~\eqref{Eq:Ham_eigen_EP}].
		(b) Eigenvectors obtained via the spectral decomposition of $\eig{3, \, 4}$ [cf. Eq.~\eqref{Eq:eigendecomposition_driven_spin}].}
	\label{fig:number_comparision}
\end{figure}

\subsection{The Liouvillian spectrum}

We now consider the complete master equation in
Eq.~\eqref{Eq:Spin_liouvillian_with_NNHEP}. The Liouvillian eigenvalues are
\begin{equation}\label{Eq:Liouvillian_spectrum_spin_system_with_HEP}
\begin{split}
\lambda_0&=0, \\
\lambda_1&=-\frac{\gamma _-}{2},\\
\lambda_{2, \, 3}&=-\frac{3}{4} \gamma_- \pm \eta/4 , \\
\end{split}
\end{equation}
while the eigenmatrices are
\begin{equation}\label{Eq:Eigenmatrices_Liouvillian_driven}
\begin{split}
\eig{0} & \propto \sss = \frac{1}{\gamma_{-}^2 + 2\omega_x^2}\begin{pmatrix}  \gamma _-^2+\omega _x^2 & i \gamma _- \omega _x \\
-i \gamma _- \omega _x & \omega _x^2
\end{pmatrix}, \\
\eig{1} & \propto
\begin{pmatrix}
0 & 1 \\
1 & 0 \\
\end{pmatrix}, \\
\eig{2, \, 3} & \propto \begin{pmatrix}
-\gamma _- \pm \eta & 4 i \omega _x \\
- 4 i \omega _x & \gamma _- \mp \eta
\end{pmatrix},
\end{split}
\end{equation}
where $\eta=\sqrt{\gamma _-^2-16\omega _x^2}$.
Hence, we expect a LEP for $\gamma_-=4\omega_x$.

In Figs.~\ref{fig:driven_Liouvillian}(a) and
\ref{fig:driven_Liouvillian}(b) we plot the real and imaginary parts of the eigenvalues $\lambda_i$ obtained in Eq.~\eqref{Eq:Liouvillian_spectrum_spin_system_with_HEP}.
Indeed, we note that, for $\gamma_-=4\omega_x$, $\lambda_{2, \, 3}$ coalesce. As expected, this EP is signaled by the coalescence of the two associated right
eigenmatrices [Fig.~\ref{fig:driven_Liouvillian}(c) and Eq.~\eqref{Eq:Eigenmatrices_Liouvillian_driven}]. We note that
the decay along the $\hat{\sigma}_x$ channel is dominated by
$\lambda_{1}$ and therefore to see interesting phenomena one
should study either $\hat{\sigma}_y$ or $\hat{\sigma}_z$.

\subsection{Comparison of HEPs and LEPs}

The question naturally arises: what is the link between the NHH
EPs and Liouvillian EPs? To answer this question we analyze the
spectra of $\eig{2}$ and $\eig{3}$ using the spectral decomposition
introduced in Sec.~\ref{Sec:Spectral_decomposition}. We obtain
\begin{equation}
\begin{split}
\eig{2, \,3} \propto \ket{\Psi_{2, \, 3}^+}\bra{\Psi_{2, \, 3}^+}
-\ket{\Psi_{2, \, 3}^-}\bra{\Psi_{2, \, 3}^-},
\end{split}
\end{equation}
where
\begin{equation}\label{Eq:eigendecomposition_driven_spin}
\begin{split}
\ket{\Psi_{2}^\pm} &\propto \left[i \left(-\gamma _-+\eta \pm
\sqrt{2 \gamma _-
	\left(\gamma _- -\eta \right)}\right), \quad 4 \omega_x \right],\\
\ket{\Psi_{3}^\pm} &\propto \left[-i \left(\gamma _-+\eta \pm
\sqrt{2 \gamma
	_- \left(\gamma _-+ \eta \right)}\right), \quad 4 \omega
_x \right].
\end{split}
\end{equation}
By comparing these with $\ket{\phi_{1}}$ and $\ket{\phi_{1}}$ in
Eq.~\eqref{Eq:Ham_eigen_EP}, we note that their structures present
several similarities when substituting $\omega_x\to 2\omega_x$. To better
capture similarity between the LEPs and HEPs, in
Fig.~\eqref{fig:number_comparision} we plot the expectation value
$\braket{\hat{\sigma}_z}$ taken over the states
$\ket{\Psi_{1, \,2}}$ [panel (a)] and $\ket{\Psi_{3, \, 4}^\pm}$
[panel (b)]. We observe that, surprisingly, the NHH captures the
behavior of $\eig{3}$, but not of $\eig{2}$, even if
$|\Re{\lambda_3}|<|\Re{\lambda_4}|$. 
Finally, we remark that the addition of the quantum jumps produces a double bifurcation, and thus we conclude that the NHH approximation is not able to capture the dynamics of $\hat{\sigma}_{x, y}$ towards the steady state.

Therefore, we may argue that the effect of quantum jumps in this
model is twofold: On the one hand, as a consequence of
Theorem~\ref{Thm:no_HEP_are_LEP}, quantum jumps modify the
structure of the eigenstates of the NHH. On the other hand,
$\{\ket{\Psi_{2}^\pm}, \, \ket{\Psi_{3}^\pm}\}$ maintain some similarities to
$\{\ket{\phi_{1}}, \, \ket{\phi_{2}}\}$. In this regard, in the next section we will see that when this two-level system is the effective description
of a bigger bosonic system in a semiclassical limit, the effect of
quantum jumps will be to introduce a mixing of the
``eigenstates'' of the corresponding NHH, according to
Theorem~\ref{Thorem:When_NHH_is_true_for_EPs}.

\begin{figure}
	\centering
	\includegraphics[width=.99\linewidth]{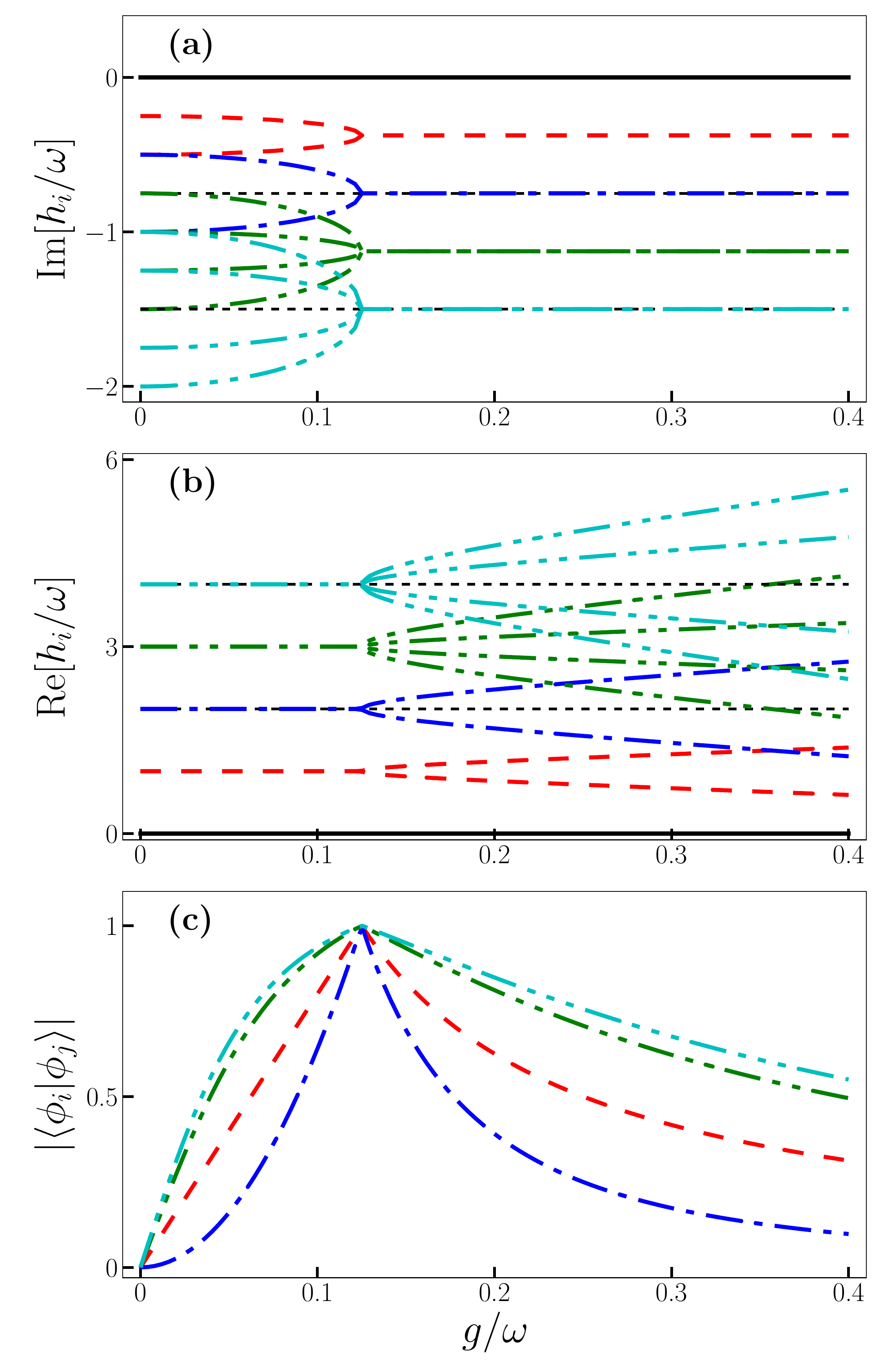}
	\caption{Spectral properties of the NHH in Eq.~\eqref{Eq:Bosonic_HEPs}.
		(a)~Imaginary and (b)~real parts of the eigenvalues as a function of $g/\omega$, i.e., the coupling between the cavities rescaled by the frequency of each cavity.	
		(c) Scalar product between the eigenvectors associated with the EPs, as a function of $g/\omega$.
		Here, the parameters used are $\gamma_a=\omega$, $\gamma_b=\omega/2$.}
	\label{fig:HEP_bosonic}
\end{figure}
\begin{figure}
	\centering
	\includegraphics[width=.99\linewidth]{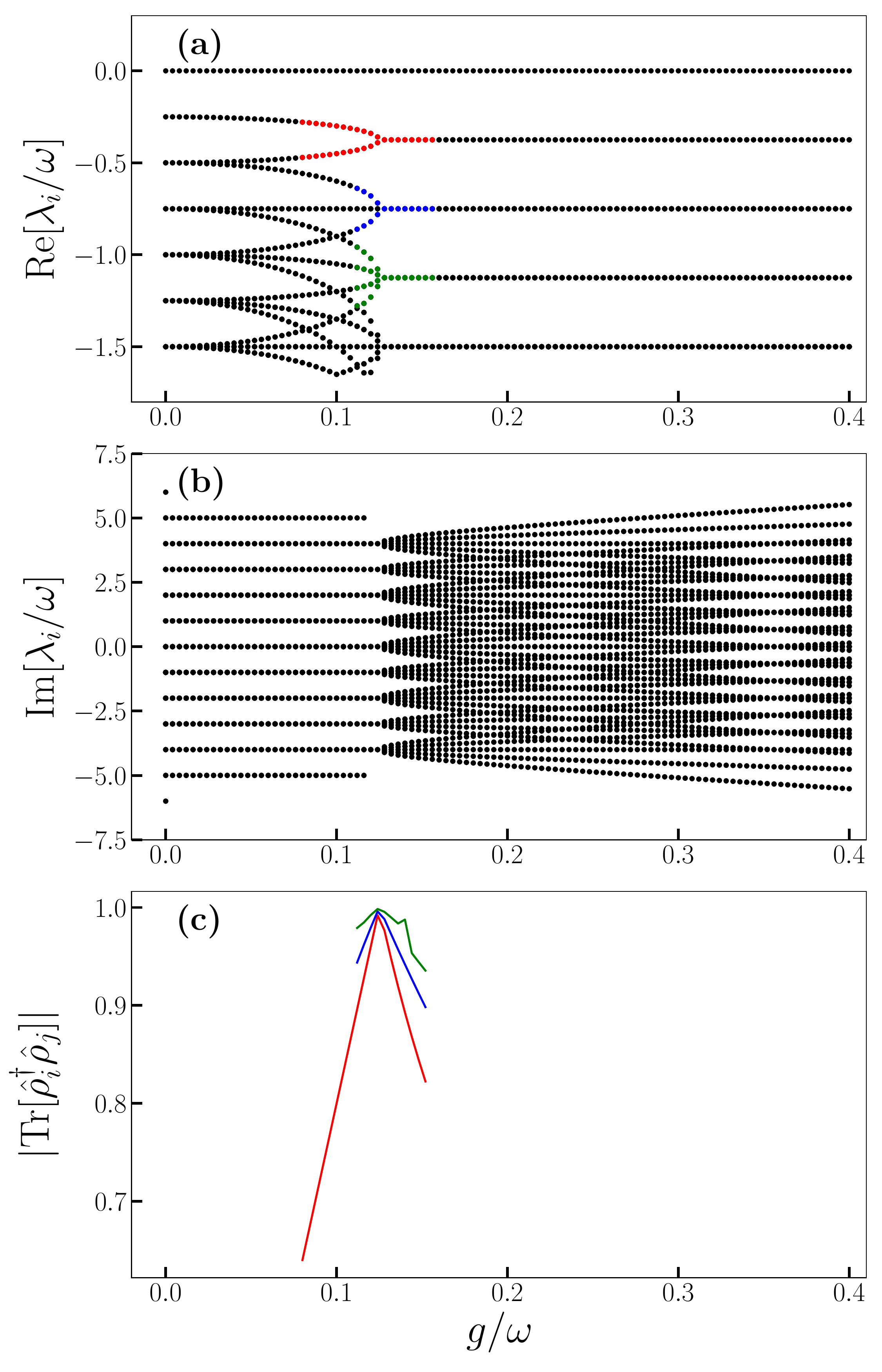}
	\caption{(Colors online) Spectral properties of the full Liouvillian in Eq.~\eqref{Eq:bosonic semiclassical_model}.
		(a) Real and (b) imaginary parts of the eigenvalues as a function of $g/\omega$.
		Note that according to Eq.~\eqref{Eq:Linblad_semiclassical}, one should compare $\Re{\lambda_i}$ with $\Im{h_i}$ in Fig.~\ref{fig:HEP_bosonic}(a), $\Im{\lambda_i}$ with $\Re{h_i}$ in Fig.~\ref{fig:HEP_bosonic}(b).	        
		(c) Scalar product between the eigenvectors associated with the EP as a function of $g/\omega$.
		We were not able to extract this quantity for all the values of $g/\omega$ due to numerical problems in correctly sorting the eigenvectors.
		Indeed, the eigenmatrices of $\LL$ no longer obey the total-excitation conservation law valid for the NHH.
		Here the parameters used are $\gamma_a=\omega$, $\gamma_b=\omega/2$.}
	\label{fig:LEP_bosonic}
\end{figure}

\section{Example of theorem 2:\\ A semiclassical model with equivalent HEPs and LEPs}
\label{Sec:Example_3} The two previous examples proved that in the
``fully quantum'' limit, the NHH fails to completely capture the
underlying physics. In this section, we consider, instead, a model
whose semiclassical limit correctly predicts the features of the
EPs as an example of
Theorem~\ref{Thorem:When_NHH_is_true_for_EPs}.

Let us consider two coupled bosonic modes, characterized by
\begin{equation}
\hat{H}= \omega \left(\hat{a}^\dagger \hat{a}+\hat{b}^\dagger \hat{b}\right) + g \left(\hat{a}^\dagger \hat{b} + \hat{b}^\dagger \hat{a} \right).
\end{equation}
and
\begin{equation}\label{Eq:bosonic semiclassical_model}
\LL = - i [\hat{H}, \bigcdot] + \frac{\gamma_a}{2} \DD[\hat{a}] + \frac{\gamma_b}{2} \DD[\hat{b}].
\end{equation}
The key element in this model is the imbalance of the dissipation
rates $\gamma_a\neq\gamma_b$, resulting in one of the two modes to
be dissipated more quickly than the other.

Physically, this model can be interpreted as photons hopping
between two cavities, one of which has a smaller quality factor
than the other. For small coupling $g$, this dissipation imbalance
tends to localize bosons in the less dissipative cavity before
eventually the system loses all the particles to the environment.
For large coupling $g$, the two modes are hybridized, and the
localization effect of dissipation cannot take place anymore. The
transition from local-to-nonlocal long-time dynamics can be
signaled by the presence of an EP.

As already shown in Ref.~\cite{Glauber1966}, the
dynamics of these two linearly coupled quantum oscillators is
(nearly) classical in nature. Note that this prototype model of a
linear coupler is mathematically equivalent to the models of a
parametric frequency converter and a beam splitter.
These models are of fundamental importance in quantum optics. In the dissipation-free case, various
two-mode phase-space quasiprobability distributions (like the
Husimi, Wigner, and Glauber-Sudarshan functions) remain constant
along purely semiclassical trajectories. Within this model, an initially semiclassical
state remains semiclassical during its evolution. In a
dissipation-free model, the degree of nonclassicality (or
classicality) of an initially quantum state remains unchanged.
This property has been used to define an operational measure of
nonclassicality~\cite{Asboth2005,Adam2015a,Adam2015b}. So, it is
convenient for us to use this model to compare the two types of
EPs in the semiclassical limit.

\subsection{Hamiltonian EPs}
The NHH associated with Eq.~\eqref{Eq:bosonic semiclassical_model}
reads
\begin{equation}\label{Eq:Bosonic_HEPs}
\hat{H}_{\rm eff}=\left( \omega- i \frac{\gamma_a}{2}\right) \hat{a}^\dagger \hat{a} + \left( \omega- i \frac{\gamma_b}{2}\right) \hat{b}^\dagger \hat{b} + g \left(\hat{a}^\dagger \hat{b} + \hat{b}^\dagger \hat{a} \right).
\end{equation}
This matrix couples subspaces with a constant number of particles,
resulting in a block-diagonal NHH.

In Figs.~\ref{fig:HEP_bosonic}(a) and \ref{fig:HEP_bosonic}(b) we
plot the imaginary and real part of the eigenvalues $h_i$ of $\hat{H}_{\rm eff}$, proving that the system admits a set of EPs, each one characterized by a
different number of excitations [cf. panel (c), where the scalar
product of the eigenvectors associated with the EPs become 1]. Most
importantly, the system exhibits all the EPs for the same value of
$g/\omega$.

\subsection{Liouvillian EPs}

We perform the same spectral analysis on the Liouvillian in
Eq.~\eqref{Eq:bosonic semiclassical_model} in
Fig.~\ref{fig:LEP_bosonic}. Similarly to
Fig.~\ref{fig:HEP_bosonic}, we observe a ladder of EPs,
characterized by really similar spectral features. We notice that,
however, in accordance to Theorem~\ref{Thm:no_HEP_are_LEP}, no EPs
occur in the steady state, even if the NHH correctly captures
the parameter $g/\omega$ for which the system has an EP.
Moreover, we notice in panel (b) that many eigenvectors have an EP with imaginary part zero, in contrast to
Fig.~\ref{fig:HEP_bosonic}(a). Finally, in panel (c) we demonstrate
that the appropriate eigenmatrices, i.e., those coalescing at the EP, have a scalar product equal to 1 for
$g=0.125 \omega$, proving that indeed the bifurcation is produced
by an EP.

\subsection{Comparison of HEPs and LEPs}

Let us begin our discussion by considering the time evolution of
the expectation value of $\braket{\hat{a}}$ and
$\braket{\hat{b}}$. For the NHH, we have $\partial_t
\braket{\hat{a}}=- i \braket{\left[\hat{a}, \hat{H}_{\rm
		eff}\right]} $, while for the Liouvillian,
\begin{equation}
\partial_t \braket{\hat{a}} =  -i  \braket{\left[\hat{a}, \hat{H}\right]} + \frac{\gamma}{2} \Tr{\hat{a} \DD[\hat{a} \rhot}.
\end{equation}
Remarkably, both equations lead to the same result,
\begin{equation}
\partial_t \begin{bmatrix}
\braket{\hat{a}} \\
\braket{\hat{b}}
\end{bmatrix} =
- i \begin{pmatrix}
\omega - i \frac{\gamma_a}{2} & g \\
g & \omega - i \frac{\gamma_b}{2}
\end{pmatrix}
\begin{bmatrix}
\braket{\hat{a}} \\
\braket{\hat{b}}
\end{bmatrix}.
\end{equation}
The previous equation confirms that the some of the HEP and LEP must be similar (i.e., same eigenvalues and eigenvectors) in order to reproduce the same dynamics of these expectation values. However, this does not mean that the whole spectral
structure is identical.

As for the similarity, we note that Theorem~\ref{Thorem:When_NHH_is_true_for_EPs} can be applied.
Indeed, let us consider the subspace with no excitation, where $\hat{H}_{\rm eff}\ket{0,0}= 0
\ket{0,0}$. It follows that ${\ket{\phi_0}=\ket{0,0}}$, since $\hat{a} \ket{0,0}=\hat{b}\ket{0,0}=0$.

We can easily verify the validity of Theorem~\ref{Thorem:When_NHH_is_true_for_EPs} by considering $\hat{H}_{\rm eff}$ in the subspace
with one excitation (i.e., $\ket{1,0}$ and $\ket{0,1}$), where
\begin{equation}
\hat{H}_{\rm eff}= \left( \omega - \frac{i \bar{\gamma} }{2} \right) \mathds{1} +  \begin{pmatrix}
- i\frac{\gamma}{2} & g \\
g &  i\frac{\gamma}{2}
\end{pmatrix},
\end{equation}
for $\bar{\gamma}=(\gamma_a + \gamma_b)/2$, and $\gamma=(\gamma_a
- \gamma_b)/2$. The eigenvalues are
\begin{equation}
h_i=\omega - \frac{i \bar{\gamma} }{2} \pm \theta,
\end{equation}
and the eigenfunctions of $\hat{H}_{\rm eff}$ are
\begin{equation}
\ket{\phi_{1,2}}=\left( -i \frac{\gamma}{2} \pm \theta\right) \ket{0,1}+ g \ket{1,0},
\end{equation}
where $\theta^2=g^2- \gamma^2/4$. 
We remark that this equation is identical to Eq.~\eqref{Eq:NHH_with_EP} by the addition of a constant term $(-i \gamma_- \mathds{1})$. 
Clearly, the model exhibits an EP for $g=\gamma/2$, where $\ket{\phi_{1}}$ and $\ket{\phi_{2}}$ coalesce.

If we consider now $\eig{1}=\ket{\phi_{1,2}}\bra{0,0}$, since we
have $\DD[\hat{a}]\eig{1}=\DD[\hat{b}]\eig{1}=0$, then
\begin{equation}
\LL \eig{1} = - i \left( \hat{H}_{\rm eff}  \eig{1}  -  \eig{1}
\hat{H}_{\rm eff}^\dagger\right)= - i h_i
\ket{\phi_{1,2}}\bra{0,0}.
\end{equation}
We have numerically confirmed this behavior for the whole spectrum of
the Liouvillian and the NHH up to a cutoff of nine excitations per site.

Finally, to correctly interpret $\eig{1}$, we consider the
eigendecomposition of $\eig{1}+\eig{2}$ (which, by construction,
is Hermitian), obtaining
$\eig{1}+\eig{2}=\ket{\Psi_1}\bra{\Psi_1}-\ket{\Psi_2}\bra{\Psi_2}$,
where
\begin{equation}
\begin{split}
\ket{\Psi_{1, \, 2}} &=\frac{\ket{0,0}}{\sqrt{2}} \pm \frac{(-1+i)\ket{0,1} + (1+i)\ket{1,0}}{2\sqrt{2}}. \\
\end{split}
\end{equation}

All the other eigenstates $\eig{i}$ which are not of the form $\ket{0,0}\bra{\phi_{l}}$ or $\ket{\phi_{l}}\bra{0,0}$, instead, have different characteristics and cannot be easily recast in terms of simple combinations of $\ket{\phi_{l}}\bra{\phi_{m}}$.

Again, we have confirmed the results of
Theorem~\ref{Thm:no_HEP_are_LEP} in an example satisfying
Theorem~\ref{Thorem:When_NHH_is_true_for_EPs}, showing that, indeed, even if
the EP position and some of the LEPs can be captured by the NHH, that is not the
case for all the eigenvectors.

\section{Conclusions and discussion}

In this paper, we addressed the question of how to define EPs in the
fully quantum regime, i.e., by including quantum jumps.
Standard EPs (i.e., HEPs) correspond to the spectra of NHHs, and
thus quantum jumps do not have any effect on these. Of course,
HEPs can be formally applied also for system far from the semiclassical limit, but the
question arises whether they properly grasp the quantum nature of
nonconservative systems.

Our proposal of defining EPs for the quantum regime is based on
analyzing the eigenfrequency and eigenstate degeneracies of the
spectra of Liouvillians. Thus, these EPs are referred to as
Liouvillian EPs or LEPs.

Our approach was motivated by the
standard Lindblad master equation and its quantum trajectory
interpretation, which includes both the continuous and the nonunitary
dissipation terms, $\hat{\Gamma}_\mu^\dagger \hat{\Gamma}_\mu \rhot
+ \rhot \hat{\Gamma}_\mu^\dagger \hat{\Gamma}_\mu$, and the
quantum jump terms, $\hat{\Gamma}_\mu \rhot
\hat{\Gamma}_\mu^\dagger$. We note that the calculation of an EP
based on NHHs includes the same continuous nonunitary dissipation
term but not the quantum jump term.

The core results of this paper concern the comparison of Liouvillian and Hamiltonian EPs.
We proved two main theorems: Theorem~1 showing that LEPs and HEPs
have essentially different properties in the quantum regime, and
Theorem~2 specifying some conditions under which LEPs and HEPs
exhibit the same properties in the semiclassical limit. We compared
explicitly LEPs and HEPs for some quantum and semiclassical
prototype models:
\begin{enumerate}[label=\textbf{Example \arabic*:}, align=left]
\item A dissipative two-level system presenting LEPs but no HEPs, i.e., a
quantum model without a semiclassical analog;
\item A dissipative two-level system presenting differences between LEPs and HEPs;
\item Two linearly
coupled dissipative quantum oscillators (a linear coupler or a
parametric frequency converter), which have semiclassical
dynamics~\cite{Glauber1966}. 
\end{enumerate}
We showed that, in general, LEPs and
HEPs can have essentially different properties (examples~1~and~2). Moreover, the model of example 3 enabled us to show explicitly
that LEPs and HEPs become essentially equivalent in the
semiclassical limit.

Note that we were not discussing here any applications of EPs.
Further research is required to generalize various semiclassical
predictions of novel photonic functionalities (mentioned in, e.g.,
reviews~\cite{Ozdemir2019,Miri2019}) to the quantum regime. These
applications might include an enhanced control of physical
processes (e.g., scattering and transmission) at EPs in composite
systems with loss, gain, and gain saturation. For example, in a
recent study of~\cite{Arkhipov2019}, the Scully-Lamb laser model
in its semiclassical limit was applied to describe the
experimentally observed light nonreciprocity and HEPs reported
in~\cite{Peng2014,Chang2014} for parity-time-symmetric
whispering-gallery microcavities. The application of the formalism
developed in this paper could enable to study LEPs in the
Scully-Lamb laser model, assuming weak gain saturation, in its full quantum regime \cite{ArkhipovarXiv19}.

Another important application of EPs could be a possible enhancement of
the sensitivity of the energy splitting and frequency detection at
HEPs, as discussed theoretically in, e.g.,
Refs.~\cite{Wiersig2014,Zhang2015,Wiersig2016,Ren2017,KuoArXiv19} and
observed experimentally in
Refs.~\cite{Chen2017,Hodaei2017,ChenNat2018,LiuPRL16}. Indeed, even a very
small perturbation applied to an NHH system at an EP can lift the
system eigenfrequency degeneracy, leading to a detectable energy
splitting. However, more detailed analyses of noise showed some
fundamental limits of HEP-enhanced
sensors~\cite{Langbein2018,Lau2018,Mortensen2018,Wolff2019,Zhang2018,Chen2019}.
In particular, a recent study~\cite{Langbein2018} indicates that
enhanced sensitivity does not necessarily imply enhanced precision
of sensors operating at EPs. 
Moreover, the importance of the unraveling protocol to obtain an enhancement of the measure sensitivity around an EP was proved in Ref.~\cite{Lau2018}.
In particular, for the nonreciprocal system of Ref.~\cite{Lau2018}, homodyne (heterodyne) detection was found to have the largest enhancement. 
In this regard, our extension to the full quantum limit allows an easy discussion of such protocols. 
Further work (which, however, is beyond the scope of this paper) is required to clarify the quantum-noise-limited performance not only of generalized LEP-based sensors, but even of standard HEP-based sensors.

LEPs can signal a second-order phase transition of driven dissipative systems, as pointed out in \cite{MingantPRA18_Spectral}. In this regard, the quest for enhanced sensitivity exploiting EPs can be corroborated by the diverging susceptibility characterizing symmetry breaking.
Extensive analyses of second-order phase transitions have been carried out for 
several types of open systems, ranging from optical cavities  \cite{BartoloPRA16,BiellaPRA17,SavonaPRA17,OverbeckPRA17,RotaPRL19}, to spin models \cite{LeePRA11,KesslerPRA12,LeePRL13,JinPRX16,RotaPRB17,RotaNJP18}, and optomechanical systems \cite{BenitoPRA16,Munuz2018}.
Moreover, EPs are relevant for the classification of topological phases of matter \cite{LeykamPRL17,GonzalesPRB17,HuPRB17,GaoPRL18,Zhou2018,LiuPRL19,BliokhNat19,MoosSciPost19,GeArXiv19}

We stress that the proposed concept of LEPs can be
applied to quantum system dynamics both with and without quantum
jumps, while the standard concept of HEPs is limited to describing
the dynamics of a system without quantum jumps. This is because
HEPs are (degenerate) eigenvalues of operators (i.e.,
non-Hermitian Hamiltonians) rather than of superoperators (e.g.,
Liouvillians).
Thus, in the semiclassical and classical regimes where quantum jumps do not
change the dynamics, the concept of LEPs is a complete alternative to the concept of HEPs.
Otherwise (i.e., when quantum jumps cannot be ignored), the approach based
on HEPs fails and should be replaced by that of, e.g., LEPs.
Moreover, the use of the Liouvillian is crucial to correctly identify and characterize LEPs without an NHH counterpart.
This formalism has the advantage to capture \emph{both} EPs resulting from the NHH and the quantum jumps, 
which otherwise could not be described in the same manner.

The analysis of EPs in the quantum regime is thus a timely
subject. These phenomena could be fully tractable in state-of-the-art
experimental platforms, such as circuit quantum-electrodynamics (QED) setups. In these
systems, the precise control of amplification, dissipation, and
coupling strength in principle allows EPs to be attained and characterized in the full quantum regime \cite{NaghilooNatPhys19,Gu2017}.

\acknowledgments{ The authors kindly acknowledge Nicola Bartolo, Alberto Biella, Naomichi Hatano, Stephen Hughes,
	Hui Jing, and {\c{S}}ahin K. \"{O}zdemir for
	insightful discussions. F.M. is supported by the FY2018 JSPS
	Postdoctoral Fellowship for Research in Japan. F.N. is supported
	in part by the: MURI Center for Dynamic Magneto-Optics via the Air
	Force Office of Scientific Research (AFOSR) (FA9550-14-1-0040),
	Army Research Office (ARO) (Grant No. W911NF-18-1-0358),
	Asian Office of Aerospace Research and Development (AOARD) (Grant
	No. FA2386-18-1-4045), Japan Science and Technology Agency (JST)
	(via the Q-LEAP program, and the CREST Grant No. JPMJCR1676),
	Japan Society for the Promotion of Science (JSPS) (JSPS-RFBR Grant
	No. 17-52-50023, and JSPS-FWO Grant No. VS.059.18N), the
	RIKEN-AIST Challenge Research Fund, and NTT Physics \& Informatics Labs.}
	\appendix

	\section{Basic properties of superoperators}
	
	\label{Sec:superoperators} Here, for pedagogical reasons and following the discussion in \cite{MingantiPHD}, we recall useful properties of superoperators, i.e., linear operators acting on the vector space of operators. That is (as stated in \cite{Carmichael_BOOK_2}), ``superoperators act on operators to
	produce new operators, just as operators act on vectors to produce new vectors.''
	
	An example of such a superoperator is the commutator
	$\mathcal{A}=\left[\hat{A}, \bigcdot\right] = \hat{A} \bigcdot -
	\bigcdot \hat{A}$. With this notation, we mean that $\mathcal{A}$
	acting on $\hat{\rho}$ is such that $\mathcal{A} \hat{\rho} =
	\hat{A} \hat{\rho} -\hat{\rho}\hat{A}$, and the dot simply
	indicates where the argument of the superoperator is to be placed.
	Moreover, we adopt the convention that the action is always on the
	operator the closest to the right-hand side of the dot. Superoperators can
	also ``embrace'' their operators, e.g., $\mathcal{A} = \hat{A}
	\bigcdot \hat{B}$ is such that $\mathcal{A}\hat{\rho} =  \hat{A}
	\hat{\rho}\hat{B} $. More generally, all superoperators can be
	represented as product of the right-hand action superoperator
	$R[\hat{O}] \, \bigcdot =\bigcdot \, \hat{O}$ and of the left-hand
	action superoperator $L[\hat{O}] \, \bigcdot=\hat{O} \, \bigcdot$.
	
	\subsection{Vectorization and matrix representation of superoperators}
	\label{App:superoperators}
	
	Since the operators form a vector space, it is possible to provide
	a vectorized representation $\vec{A}$ of each element $\hat{A}$ in
	$H \otimes H$. For example,
	\begin{equation}
	\hat{A}=\begin{pmatrix}
	a & b \\
	c & d
	\end{pmatrix} \longrightarrow \vec{A} = \begin{pmatrix}
	a \\ b \\ c \\ d
	\end{pmatrix}.
	\end{equation}
	Consequently, to any \emph{linear} superoperator $\mathcal{A}$ it
	is possible to associate its matrix representation
	$\bar{\bar{\mathcal{A}}}$.
	
	More generally, given an orthonormal basis of the Hilbert space
	$\{\ket{n}\}$, for a generic operator $\hat{\xi}$ we have
	\begin{equation}\label{Eq:vectorisation}
	\begin{split}
	\hat{\xi}=\sum_{m,n} c_{m,n} \ket{m}\bra{n}\longrightarrow
	\vec{\xi}  &=\sum_{m,n} c_{m,n} \ket{m}\otimes \bra{n}^{\rm TR}
	\\ &=\sum_{m,n} c_{m,n} \ket{m}\otimes \ket{n^*},
	\end{split}
	\end{equation}
	where $\rm TR$ represents the transpose, while $*$ is the
	complex conjugate. To obtain the matrix form of any superoperator,
	we must describe the right- and left-hand action superoperators
	$R[\hat{O}]$ and $L[\hat{O}]$ as matrices $\supmat{R}[\hat{O}]$
	and $\supmat{L}[\hat{O}]$. One has
	\begin{equation}\label{Eq:RightAction}
	\begin{split}
	\supmat{R}[\hat{O}] \vec{\xi} & =\supmat{R}[\hat{O}] \sum_{m,n}
	c_{m,n} \ket{m}\otimes \ket{n^*}= \overrightarrow{\hat{\xi}
		\hat{O}}
	\\ &= \sum_{m,n} c_{m,n}  \, \ket{m}\otimes (\bra{n} \hat{O} )^{\rm TR}
	\\ &=\sum_{m,n} c_{m,n}  \, \ket{m}\otimes ( \hat{O}^{\rm TR}\ket{n ^*}) = (\mathds{1}\otimes \hat{O}^{\rm TR}) \vec{\xi}.
	\end{split}
	\end{equation}
	Analogously, we have
	\begin{equation}\label{Eq:LeftAction}
	\supmat{L}[\hat{O}] \vec{\xi}= (\hat{O}\otimes \mathds{1})
	\vec{\xi}.
	\end{equation}
	
	From the result of
	Eqs.~\eqref{Eq:RightAction}~and~\eqref{Eq:LeftAction}, we can
	eventually write any Liouvillian $\LL= - i \left[\hat{H},
	\bigcdot\right] +  \mathcal{D}[\hat{\Gamma}]$  (for simplicity,
	here, with only one jump operator $\hat{\Gamma}$) in the form
	\begin{equation}
	\begin{split}
	\supmat{\mathcal{L}}&= -i \left[\supmat{L}(\hat{H}) -
	\supmat{R}(\hat{H}) \right]+\supmat{L}(\hat{\Gamma})
	\supmat{R}(\hat{\Gamma}^\dagger) -
	\frac{\supmat{L}(\hat{\Gamma}^\dagger \hat{\Gamma})}{2}
	-\frac{\supmat{R}(\hat{\Gamma}^\dagger \hat{\Gamma})}{2} \\ & = -i
	\left(\hat{H}\otimes \mathds{1}  - \mathds{1}\otimes \hat{H}^{\rm
		TR}\right) +  \hat{\Gamma}\otimes
	\hat{\Gamma}^*-\frac{\hat{\Gamma}^\dagger \hat{\Gamma}\otimes
		\mathds{1}}{2} -  \frac{\mathds{1} \otimes  \hat{\Gamma}^{\rm TR}
		\hat{\Gamma}^*}{2}.
	\end{split}
	\end{equation}
	Similarly, we obtain the Liouvillian without quantum jumps $\LL'$
	and its matrix representation
	\begin{equation}
	\supmat{\mathcal{L}'} = -i \left(\hat{H}\otimes \mathds{1}  - \mathds{1}\otimes \hat{H}^{\rm TR}\right) - \frac{\hat{\Gamma}^\dagger \hat{\Gamma}\otimes \mathds{1}+  \mathds{1} \otimes  \hat{\Gamma}^{\rm TR} \hat{\Gamma}^*}{2},
	\end{equation}
	while the quantum jump term reads $\mathcal{J}[\hat{\Gamma}]
	=\hat{\Gamma}\otimes \hat{\Gamma}^*$.
	
	This procedure can be easily generalized to multiple quantum jump
	operators.

	\subsection{Hilbert-Schmidt inner product and Hermitian conjugation}
	
	In order to discuss the coalescence of eigenstates at an EP, it
	is useful to introduce a scalar product. Since there is no
	intrinsic definition of an inner product in the operator space $H
	\otimes H$, we introduce the Hilbert-Schmidt product:
	\begin{equation}
	\langle\hat{A}\vert \hat{B}\rangle = \Tr{\hat{A}^\dagger \hat{B}}.
	\end{equation}
	Hence, the norm of an operator is
	\begin{equation}
	\|\hat{A}\|^2=\Tr{\hat{A}^\dagger \hat{A}}.
	\end{equation}
	That is, given two matrices
	\begin{equation}
	\hat{A}  =\begin{pmatrix}
	a & b \\
	c & d
	\end{pmatrix}, \quad    \hat{E}=\begin{pmatrix}
	e & f \\
	g & h
	\end{pmatrix},
	\end{equation}
	one has
	\begin{equation}
	\begin{split}
	\braket{\hat{A}|\hat{E}} & = \begin{pmatrix} a^* & b^* & c^* & d^*
	\end{pmatrix} \begin{pmatrix}
	e & f & g & h
	\end{pmatrix}^{\rm TR} \\ & =a^* e + b^* f + c^* g + d^*h =\Tr{\hat{A}^\dagger \hat{E}}.
	\end{split}
	\end{equation}
	Most importantly, having introduced the inner product for the
	operators, it is possible to introduce the Hermitian adjoint
	\footnote{There are several different notations in literature to
		indicate Hermitian conjugation, and the symbol $\dagger$ is used
		with different meanings. In particular, in Ref.~\cite{Carmichael_BOOK_2} the symbol $\hat{A}^\dagger$ 		indicates a conjugate ``associated'' superoperator.} of
	$\mathcal{A}$, which by definition is $\mathcal{A}^\dagger$ such
	that:
	\begin{equation}
	\braket{\hat{\xi} | \mathcal{A} \hat{\chi}} = \braket{\mathcal{A}^\dagger \hat{\xi} |  \hat{\chi}}.
	\end{equation}
	The rules to obtain such an adjoin, however, \emph{are not} the same
	as in the case of operators. Consider the most general linear
	superoperator $\mathcal{A} = \hat{A} \bigcdot \hat{B}$. Exploiting
	the definition of the Hermitian adjoin we have
	\begin{equation}
	\begin{split}
	\braket{\hat{\xi} | \mathcal{A} \hat{\chi}} &=\Tr{\xi^\dagger
		\hat{A} \hat{\chi} \hat{B}}= \Tr{\hat{B} \xi^\dagger \hat{A}
		\hat{\chi}}=
	\Tr{(\hat{A}^\dagger \hat{\xi} \hat{B}^\dagger)^\dagger \hat{\chi}} \\
	&= \Tr{(\mathcal{A}^\dagger \hat{\xi})^\dagger \hat{\chi}}=
	\braket{\mathcal{A}^\dagger \hat{\xi} | \hat{\chi}}.
	\end{split}
	\end{equation}
	We conclude that
	\begin{equation}
	\mathcal{A}^\dagger = \hat{A}^\dagger \bigcdot \hat{B}^\dagger.
	\end{equation}
	Note that
	\begin{equation}
	\left(\mathcal{A} \hat{\xi}\right)^\dagger = \left(\hat{A}
	\hat{\xi} \hat{B}\right)^\dagger = \hat{B}^\dagger
	\hat{\xi}^\dagger \hat{A}^\dagger \neq \mathcal{A}^\dagger
	\hat{\xi}^\dagger.
	\end{equation}



%

\end{document}